\documentclass[preprint2]{aastex}
\usepackage{color,graphicx,times}

\slugcomment{Pre-Print}
\shorttitle{ Identification of
Disease in  Complex Pedigrees}
\shortauthors{Leibon, et al.}

\begin{document}

\title{A simple computational method for the identification of
disease-associated loci in complex, incomplete pedigrees}

\author{Gregory Leibon,\altaffilmark{1,5} 
Dan Rockmore,\altaffilmark{1,2,5} and,  Martin R. Pollak\altaffilmark{3, 4}}

\altaffiltext{1}{Department of Mathematics, Dartmouth College, Hanover, NH 03755}
\altaffiltext{2}{Department of Computer Science, Dartmouth College, Hanover, NH 03755}
\altaffiltext{3}{Department of Medicine, Brigham and Women's Hospital and Harvard Medical School,
4 Blackfan Circle, Boston, MA 02115}
\altaffiltext{4} {Supported in part by NIH grant DK54931.}
\altaffiltext{5} {Supported in part by NIH grant GM075310.}

\begin{abstract}

We present an approach, called the {\it Shadow Method}, for the identification of disease loci from dense genetic marker maps in complex, potentially incomplete pedigrees. 
{\it Shadow} is a simple method based on an analysis of the patterns of obligate meiotic recombination events in genotypic data.
This method  can be applied to any high density marker map and was specifically designed  to exploit  the fact that extremely dense marker maps  are  becoming more readily available. We also describe how to interpret and associate meaningful  $P$-Values to the results.  {Shadow} has significant advantages over traditional parametric linkage analysis methods in that it can be readily applied even in cases in which the topology of a pedigree or pedigrees
 can only be partially determined.   In addition,  {Shadow} is robust to variability in a range of parameters and in particular does not require prior knowledge of mode of inheritance, penetrance or clinical misdiagnosis rate.  {Shadow} can be used for any SNP data, but   is especially effective when applied to  dense samplings.  Our primary example uses data from Affymetrix 100k SNPChip samples in which we illustrate our approach by analyzing simulated data as well as genome-wide SNP data from two pedigrees with inherited forms of kidney failure, one of which is compared with a typical LOD score analysis.  
\end{abstract}


\maketitle

\section{Introduction}

Studies of genetic disease have been remarkably successful in identifying disease genes and novel biological pathways.  For family-based analyses of phenotypes with single, highly penetrant disease alleles, the first step is the identification of a locus harboring the mutant allele. This requires the acquisition and subsequent analysis of a significant amount of genetic data. As regards the former, the ease with which investigators can accomplish genome-wide genotyping has increased tremendously in recent years.  For example, one commercial microarray technology (Affymetrix SNPChip) now allows rapid chip-based genotyping of approximately $10,000, 100,000,$ and $500,000$ SNPs (see \citet{Mat1} and \cite{Mat2}).  

Most of the currently available linkage approaches were originally developed with the goal of extracting as much information as possible from a relatively small set of markers.  
We base our approach on the fact that with very dense genetic maps, we can ignore markers that are not fully informative and still extract most of the useful genetic information.  In essence, our method is based on identifying obligate recombination events and using the distribution of these events to identify genomic regions inherited identical by descent (IBD).  This allows us to handle the complicated requirements of real data and the often complex and incompletely known structures of available pedigrees. We call our technique the {Shadow Method} and introduce it in the next section.

Our motivation for the development of {Shadow}  is severalfold. Perhaps most important is the fact that  available software is overmatched by the great number of computations required in order to calculate parametric or non-parametric 
LOD scores for large pedigrees and large data sets.  It  is known that using standard methods, the size of the calculation (as measured in the number of arithmetic operations)   increases exponentially in pedigree size  or number of markers used (the various elaborations of the  Elston-Stewart algorithm as in  \citet{Ott} 
and  the NPL algorithm as  \citet{Kruglyak} respectively).   In contrast, the computational load of  {Shadow} only  grows   linearly  with the number of markers and at a rate that is less than exponential  in pedigree size.  In the worst  case scenario,  it  increases exponentially in  {\em sample}\footnote{ In this paper we draw the distinction between the pedigree members and the  {\em samples}, the latter of which are those people in the pedigree for whom we possess a  genotyped DNA sample.} size,  but is   independent of pedigree size.  This enables us to analyze 
large pedigrees.    

Computational complexity is just one concern. We are also cognizant of the fact that  in analyses of large complex pedigrees, it can be extremely useful for  investigators to have an index of which regions are most likely to harbor disease genes by virtue of the  of
 sharing regions IBD in affected individuals, as well as a measure, given data from a subset of a pedigree, of distance from IBD for any region of the genome.  This relies on the computation of something we call the {Shadow} function at the locus $x$, denoted $S(x)$. 
It is effectively a measure of just how inconsistent the data is with the hypothesis that the pattern of inheritance at a given locus is from IBD. In particular,  $S(x) = 0$ implies  IBD at $x$.

Thus, the {Shadow} method  is a conceptually and computationally simple technique with several features that we believe make it useful for the analysis of large, complex, and perhaps incomplete pedigrees, particularly for relatively rare diseases caused by uncommon genetic variants of large effect:  (1)  {Shadow} enables rapid identification of genetic regions most likely to harbor IBD regions in pedigrees; (2) {Shadow} measures
how inconsistent such regions (and in fact all regions) are from being IBD; and (3) {Shadow} helps to identify the source of such inconsistencies in ``almost IBD'' regions.    We also develop methods to assess how likely we are to find such IBD or ``almost IBD'' regions by chance.   The specifics of this measure and the details of its interpretation are presented in the next section.

We illustrate the use of {Shadow} by analyzing both simulated data as well as genome-wide SNP data from two pedigrees with inherited forms of kidney disease.  The pedigrees are illustrated in Figure~\ref{fig:pedigrees}. The family  FS-Z has a relatively simple pedigree and it is known that  the responsible gene defect is a point mutation in the TRPC6 gene on chromosome 11q (\citet{Reiser}).  In this case a  full multi-point linkage analysis will work well and we compare our results to a LOD score analysis.  The second family we analyze, the FG-FM family, has an incomplete and large pedigree, a situation which makes standard linkage approaches unreliable and/or impossible.

\section{The Shadow Function - Measuring distance from IBD}\label{method}

\subsection{Definition of Shadow function}

At the core of the {Shadow} method is the idea that  the sample data provides us with a means to measure for each locus $x$ the degree to which the data is {\bf in}consistent with the hypothesis that the region around $x$ is IBD and thus is possibly within a disease-harboring allele. We call this  measure the {Shadow} function and denote it as  $S$. Since we focus  on inconsistency, a locus $x$  that is consistent with the IBD assumption has $S(x)= 0$, reflecting that it is distance $0$ from being IBD. 

To articulate this distance we use the familiar notion of an {\em inheritance vector}, as introduced in \citet{Kruglyak}.  
 Recall, an inheritance vector  $v$ is a vector of ones and zeros that tells us which copy of a marker  is passed on during a particular meiosis process in our pedigree.  In particular, if we label one of the chromosomes in each homologous chromosomal pair with a zero and the other with a one, then we have an inheritance vector $v(x)$ defined at each locus $x$. 
   The value of 
$S$ approximates the minimal number of changes in the inheritance vector  necessary for $v(x)$  to be  ideally consistent with a disease allele being located at that point $x$.  Since our examples use only affected samples, this gives us an  estimate  of  the minimal number of changes in the inheritance vector  necessary for  inheritance vector at $x$ to be IBD.  In Section~\ref{Comp} we explain how to include controls. 


The exact sense of distance is captured by the following definition:

{\it 
{\bf Definition:}   
For a given inheritance vector $v$, let $m(v)$ denote the minimal number of changes (bit-flips) necessary to make the vector IBD.  We call a partition of our samples\footnote{A partition of the set of samples is simply its decomposition into a collection of disjoint subsets.} {\it consistent} with a given inheritance vector 
$v$ if the samples in each part of the partition are IBD from some common founder using $v$. We let $Part(v)$ be the set of  the partitions consistent with $v$.   Similarly we denote as   $Inh(P)$ the  set of inheritance vectors 
with which $P$ is  consistent.  Then we define $S$ to be
 \[ S(x) =  \min_{ P \in Part(v(x)) }  \left(  \min_{v  \in Inh(P)}  m(v) \right).  \] 
}

For example, for a simple pedigree with autosomal dominant inheritance and $0\%$ phenocopy rate,  $S(\mbox{disease  locus})=0$.  

Figure~\ref{Increase} gives us a first illustration of the function  $S$.  The {Shadow} in  Figure~\ref{Increase} was constructed from a simulated FS-Z family assumed  to have the disease at $1$ morgan from the p end of the chromosome 11. That is, we ran twenty simulations of allele segregation in chromosome 11 consistent with the pedigree for FS-Z and a disease locus  at the  TRPC6 locus and chose  two IBD regions for illustrative purposes.   Hence $S(\mbox{Ch 11},\mbox{1 morgan})=0$ since this location is fully consistent with harboring a disease allele. Each time a crossover occurs in meiosis, there is a change in the inheritance vector.  There are crossovers on both sides of the disease locus, and 
as we move from the disease locus past  such a crossover the value of $S$ goes from $0$ to $1$.   In general, the ``corners'' of the {Shadow} curve (see Figure~\ref{Increase}) will represent crossovers that have had an effect on what our data will look like from the point of view of our samples.   Notice that  we have used a convention where the {\it distance}-axis ($y$-axis) has its minimal value of $0$ at the top and  increases as we move down the axis.  

In real data, at the disease locus $x$  the {Shadow} may have $S>0$.  In such a case,  the value of $S$ is  easy to interpret. Namely 
\[     S(x) =\#   \mbox{of inconsistencies}  \]
 where  an   {\it inconsistency} may be either an unanticipated founder  or a person who 
has an indistinguishable phenotype but not a disease allele (i.e., a phenocopy).

\subsubsection{How to use $S$ - the $P$-Value} 

\begin{figure}
    \begin{minipage}[t]{8cm}
 \includegraphics[width=0.75 \textwidth]{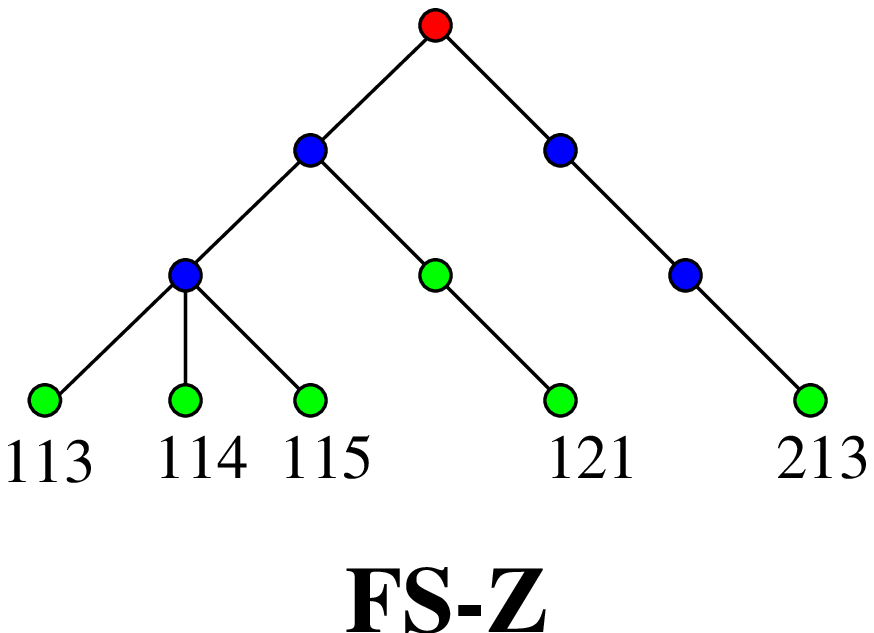}
 \end{minipage}
 \begin{minipage}[t]{11cm}
 \includegraphics[width=0.75 \textwidth]{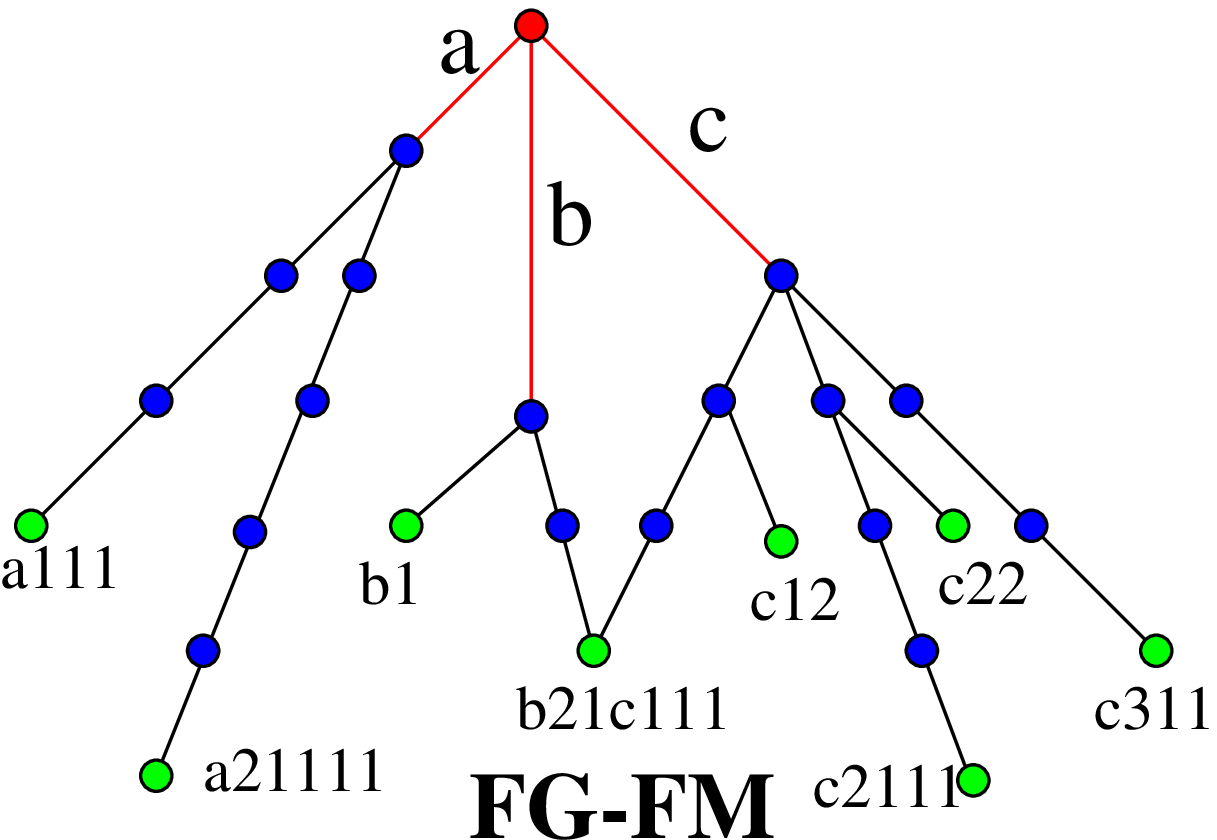}
  \end{minipage}
\caption{\label{fig:pedigrees} Examples of collapsed pedigrees of the families analyzed here.  A {\it collapsed pedigree} only includes people for whom there exist a genotyped DNA sample (in green), the non-founders (in blue), and founders that in the ideal disease associated scenario contributed a disease allele (in red).  Red edges are hypothetical and  $a$, $b$, and $c$ represent the number of non-founders along the hypothetical edge.  For the FG-FM family this is a minimally complicated pedigree consistent with a unique ``red'' founder and in our analysis we assume $a=c=1$ and $b=2$.
    }\label{FMFZ}  
\end{figure}

The use of $S$ is very similar to the use of the LOD score function $LOD(x)$. Figure~\ref{LodComp} compares the two. In particular,
 if we knew $S$ but not the disease locus (or loci),  then we would identify the 
region(s) in the genome where $S$ is minimal are likely candidates. The next step in the evaluation of such regions is 
to determine how likely it is that such a scenario is  the result of chance alone.  We call the probability of this scenario being due to chance alone the event's {\it $P$-Value}.    If the $P$-Value is small, then we can conclude with some specific computed probability that the  certainty that a disease locus is in this region and interpret the value of $S$ at this point as the number of inconsistencies.  As with the LOD score method (or any method) if this $P$-Value is large then it will be difficult to distinguish  a disease allele-harboring region from a  chance IBD region and this will  lead to a high false positive rate. In Section ~\ref{PVal} we review the process of estimating the $P$-Value. For example, if we make the definitions 
  \[ ch_i = \mbox{Morgan length of  chromosome} i \] 
  and 
   \[ B  = \#(\mbox{Branches in collapsed pedigree}) \]    
  for a tree  we estimate 
\[ P  <   4 \sum_{i=1}^{22} \frac{B \cdot ch_i+1}{2^B}.  \] 
  In general, the  size and complexity of the pedigree will influence the $P$-Value and hence the number of inconsistencies that can exist at a true disease locus in the given family before  this method will give  false positives. Consider the pedigrees  in Figure~\ref{FMFZ}.  We find that in the FS-Z pedigree, the presence of any inconsistency will be fatal.  By contrast, in the FG-FM family a single inconsistency would still yield significant results.  Specifically, a region where $S=1$ could still be regarded as likely to harbor a disease allele, given the existence of one inconsistency as defined above.

\subsection{Approximation of the  {Shadow} function}

  In practice we do not have access to the actual {Shadow}, but an approximation  denoted  as $S_M$.
   The idea behind this approximation is very simple, namely that we can identify obligate recombination events between two individuals
if they have incompatible alleles at some marker.  With our SNP data   
 if we see that Person 1 has alleles AA at the same SNP
locus where Person 2 has alleles BB then we have an obligate recombination event.  With a very dense and polymorphic SNP map, we can be reasonably sure that if a sufficiently
long consecutive stretch of SNPs occurs without such an obligate
recombination event, then these individuals share (at least) 1
chromosomal region identical by descent.   We say that such a streak of markers is {\it consistent} with a partition $P$ (of samples) if each part of the partition contains no obligate recombination events throughout the streak.      


We will view the streak as a non-coincidence if it exceed the critical length of $M$ markers  (how to choose $M$ is  explored at length in Section~\ref{Estimates}). 


   {\it 
     {\bf Definition:}    Let  $Part_M(x)$  be the set of  partitions with the property that there exists a streak of length at least $M$ and  containing $x$ that  
   is  consistent with this partition.  Then we define $S_M$ to be 
   \[ S_M(x) = \min_{ P \in Part_M(x) }  \left(  \min_{v  \in Inh(P)}  m(v) \right).  \] 
   }

\begin{figure}
 \begin{minipage}[t]{3cm}
 \includegraphics[width=0.9\textwidth]{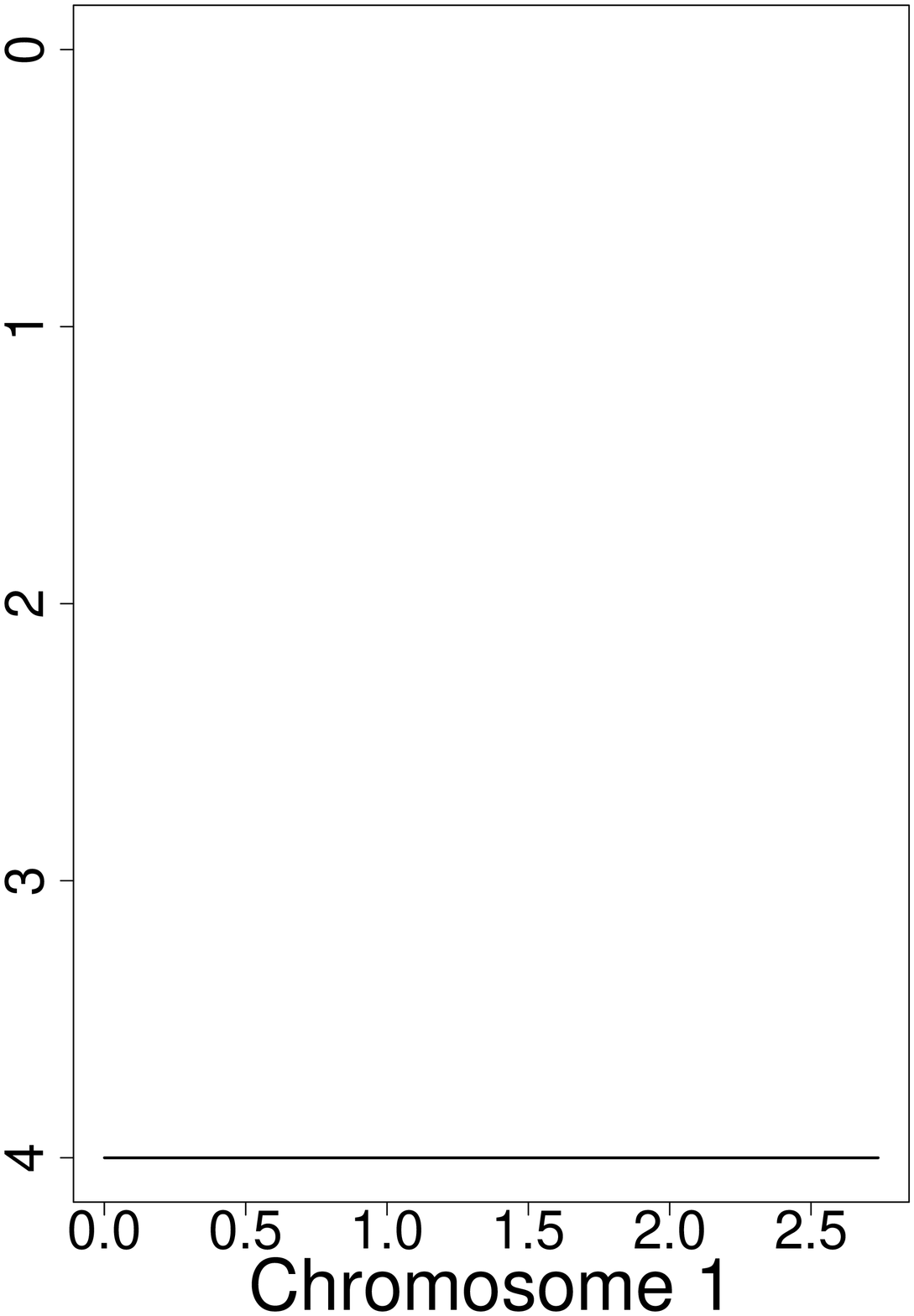} 
 \end{minipage}
 \hfill
 \begin{minipage}[t]{3cm}
 \includegraphics[width=0.9 \textwidth]{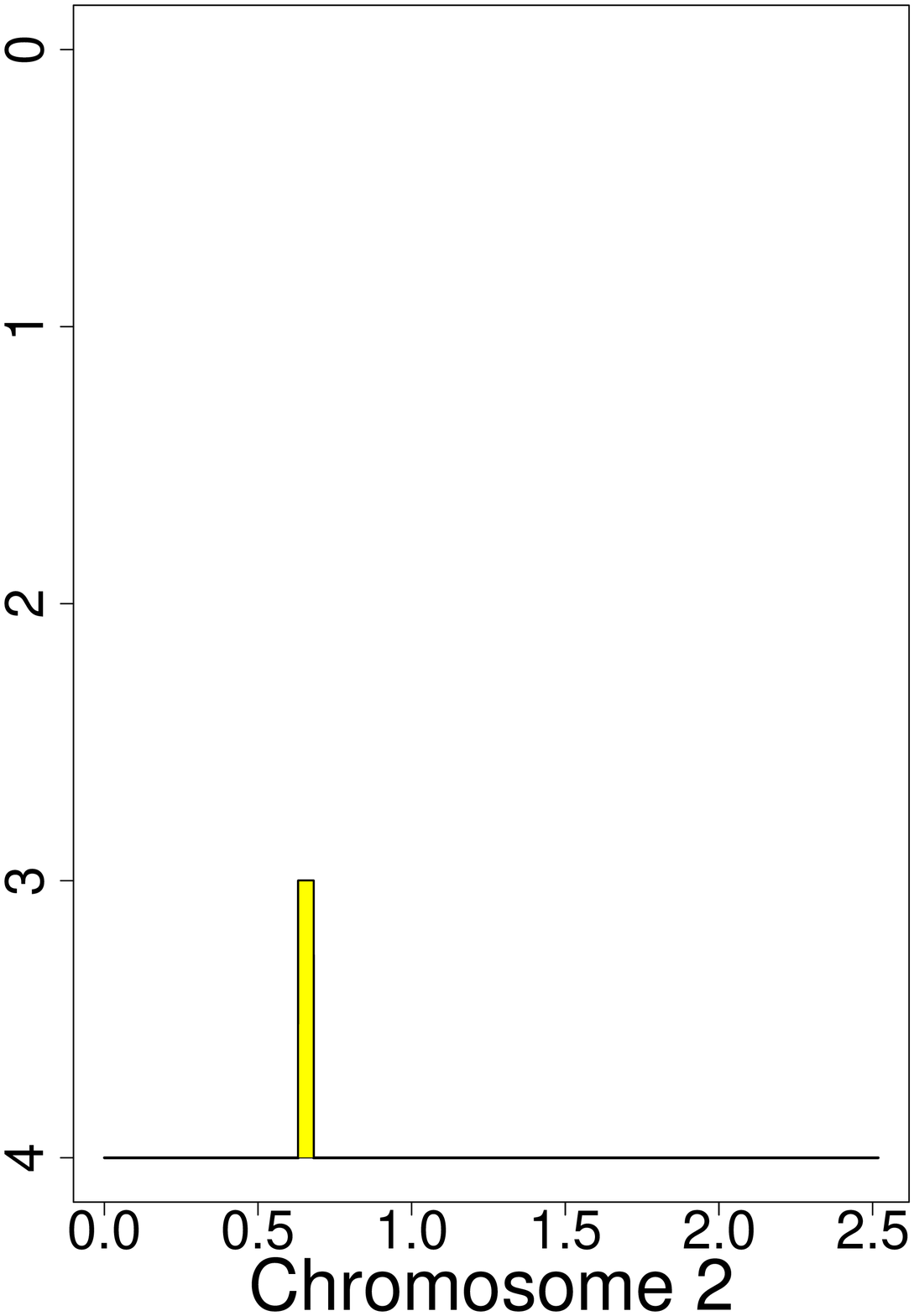}
 \end{minipage}
 \hfill
 \begin{minipage}[t]{3cm}
 \includegraphics[width=0.9 \textwidth]{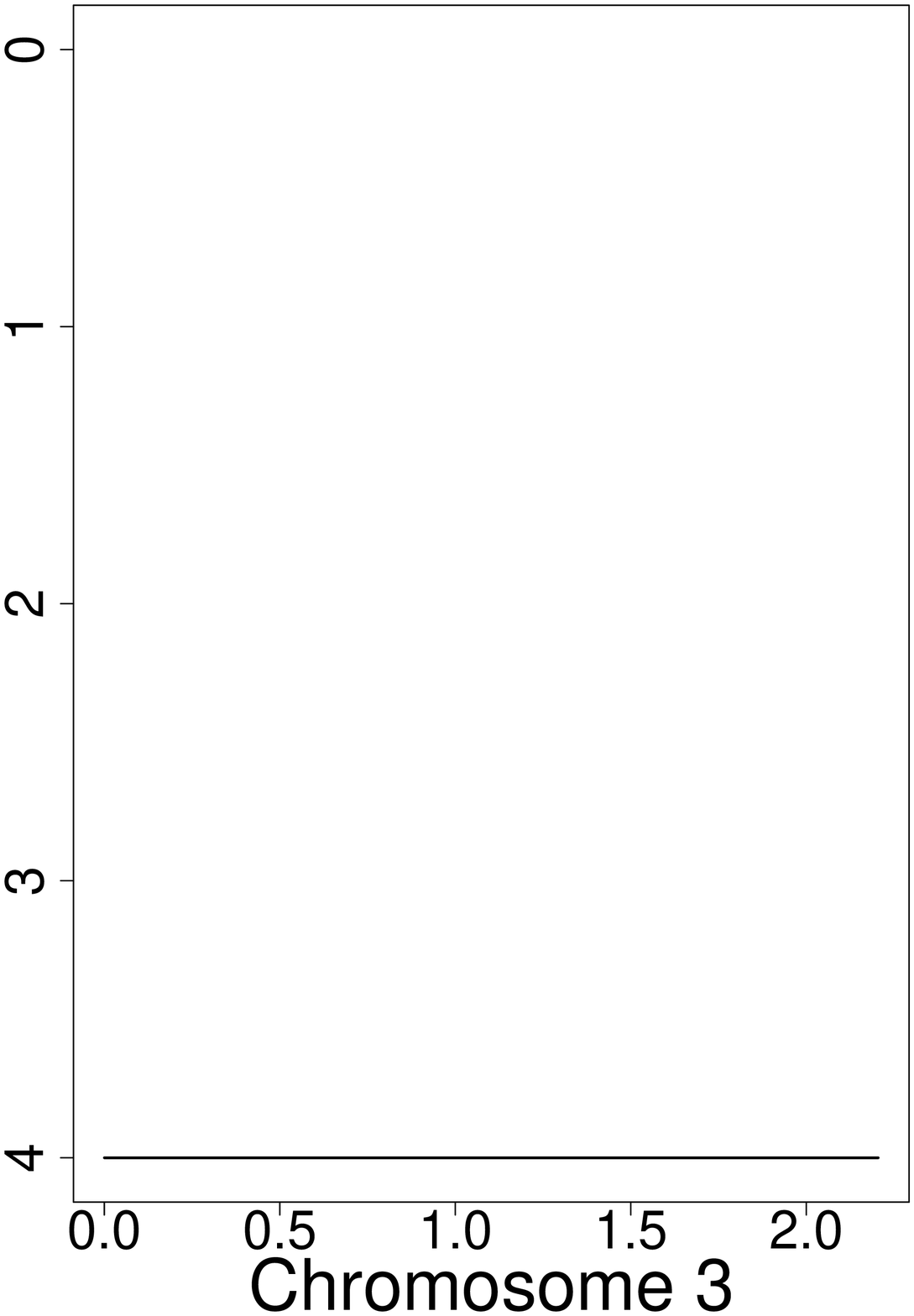}
  \end{minipage}
 \hfill
 \begin{minipage}[t]{3cm}
 \includegraphics[width=0.9 \textwidth]{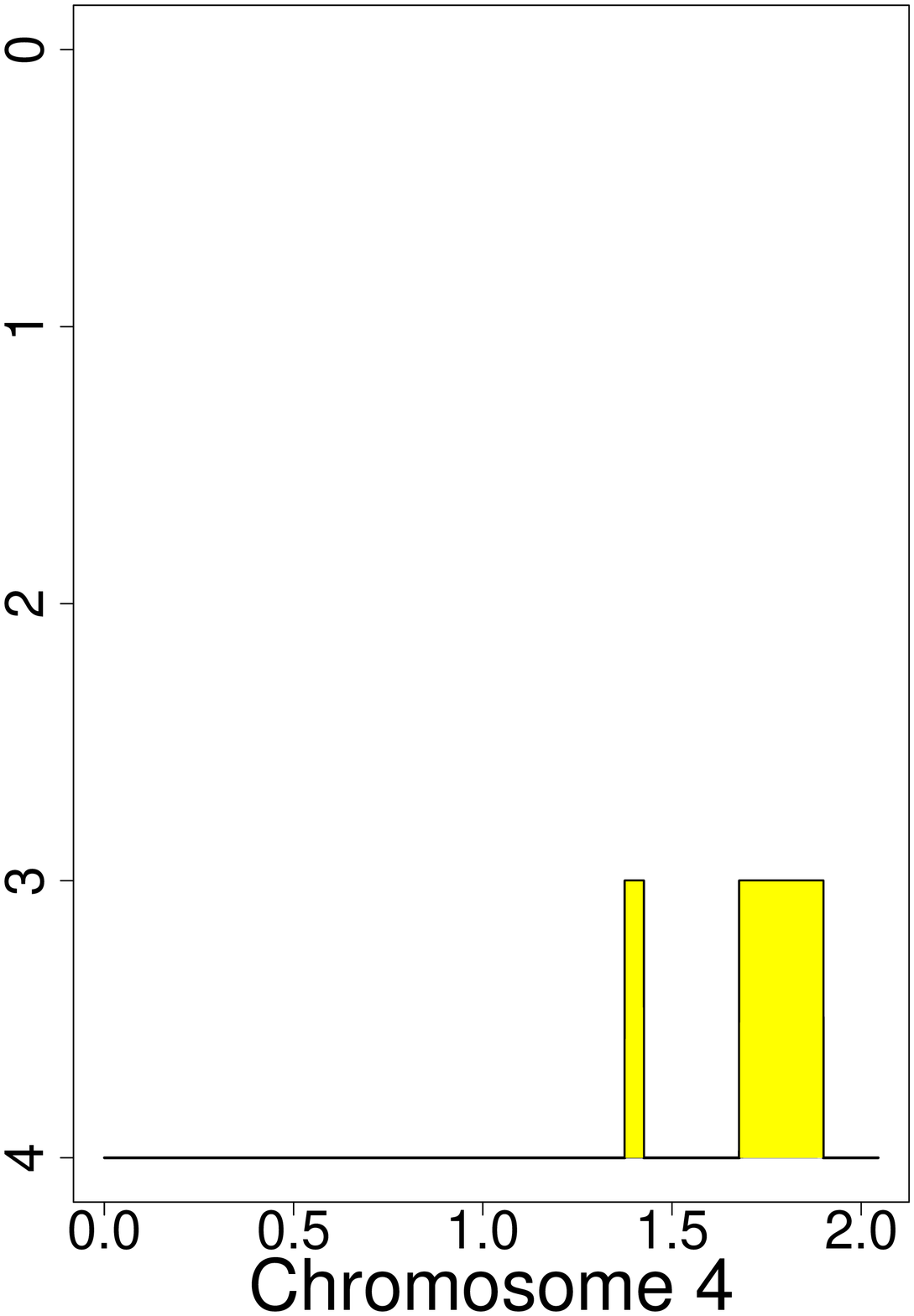}
 \end{minipage}
 \hfill
 \begin{minipage}[t]{3cm}
 \includegraphics[width=0.9\textwidth]{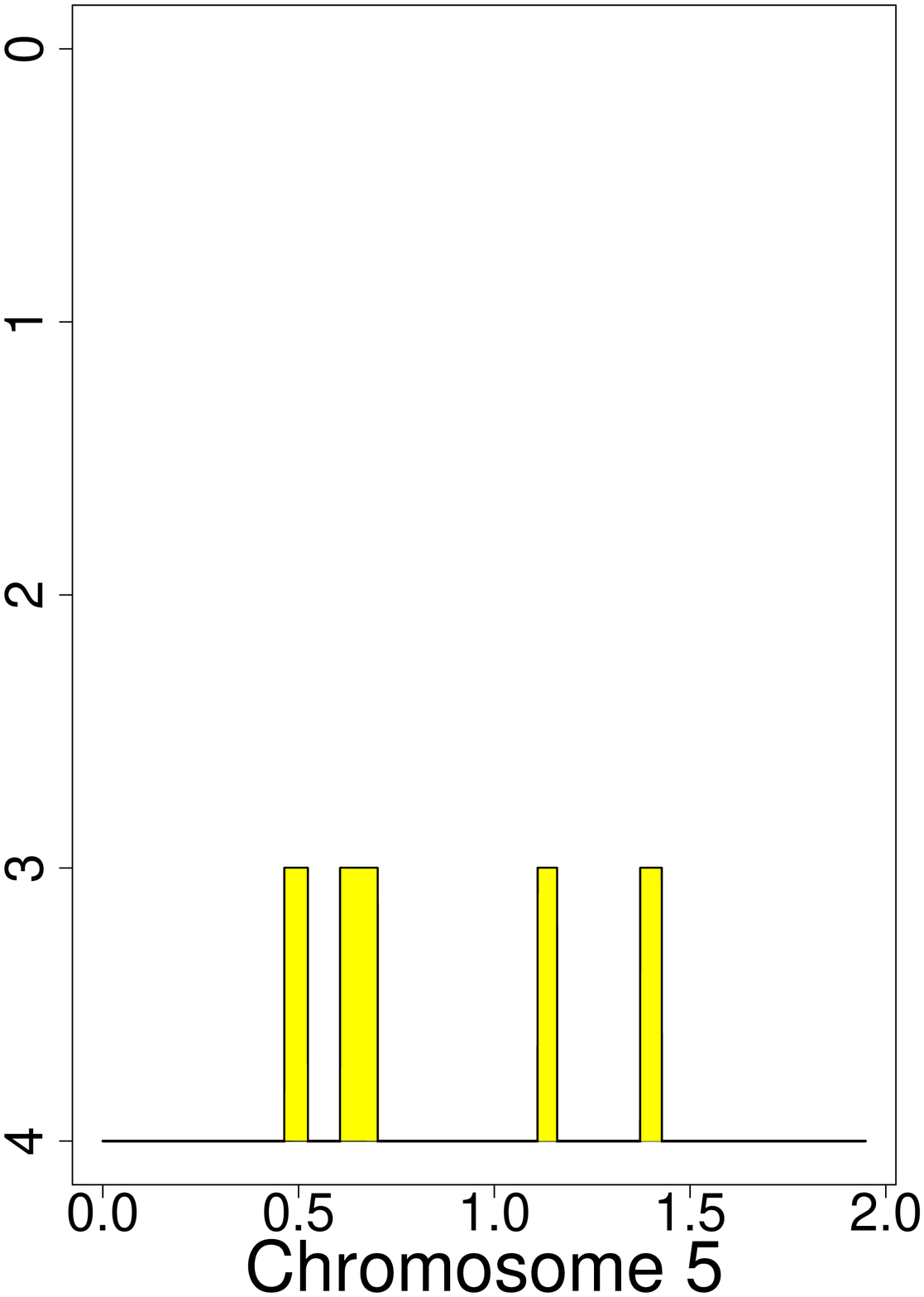} 
 \end{minipage}
 \hfill
 \begin{minipage}[t]{3cm}
 \includegraphics[width=0.9 \textwidth]{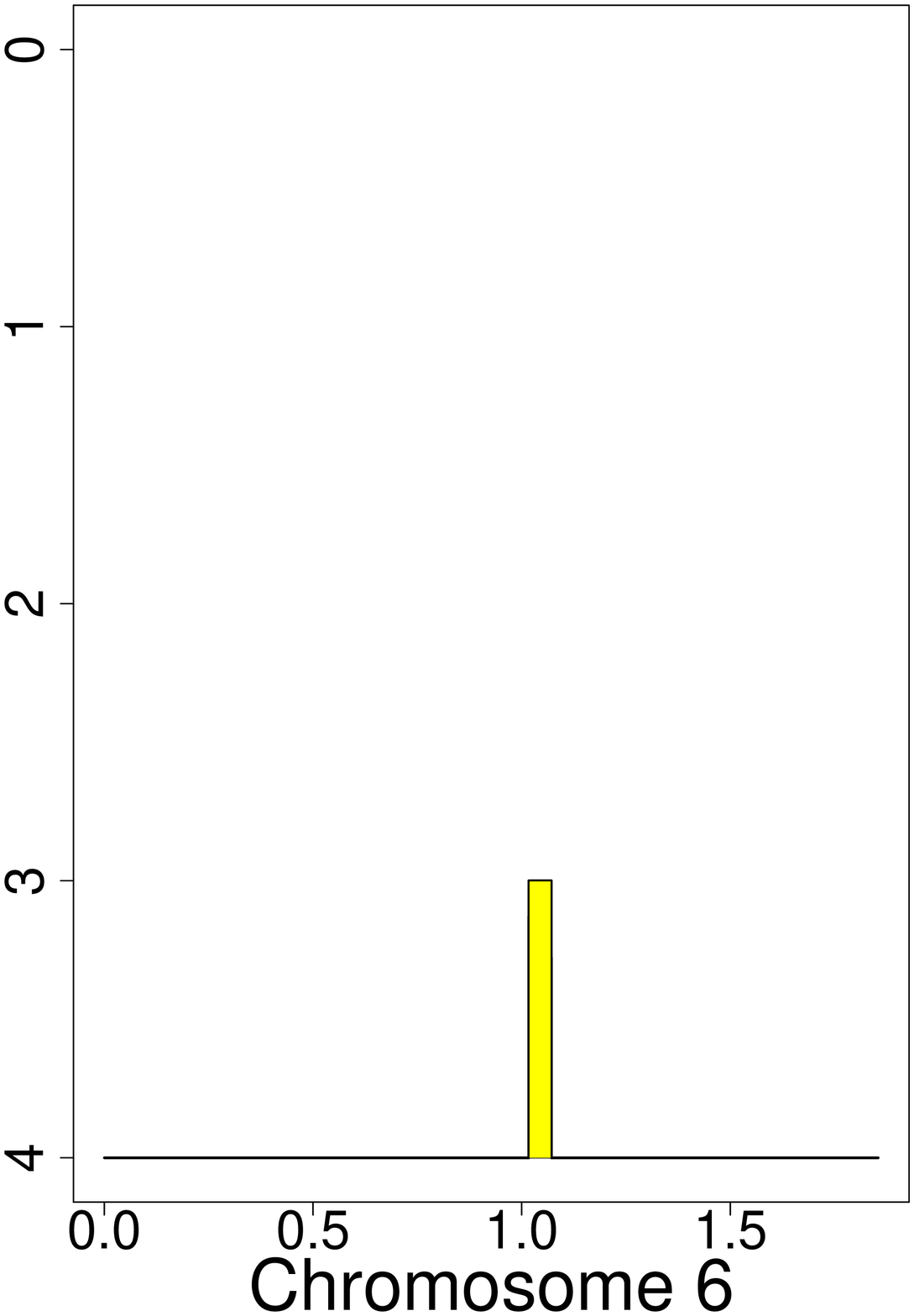}
 \end{minipage}
 \hfill
 \begin{minipage}[t]{3cm}
 \includegraphics[width=0.9 \textwidth]{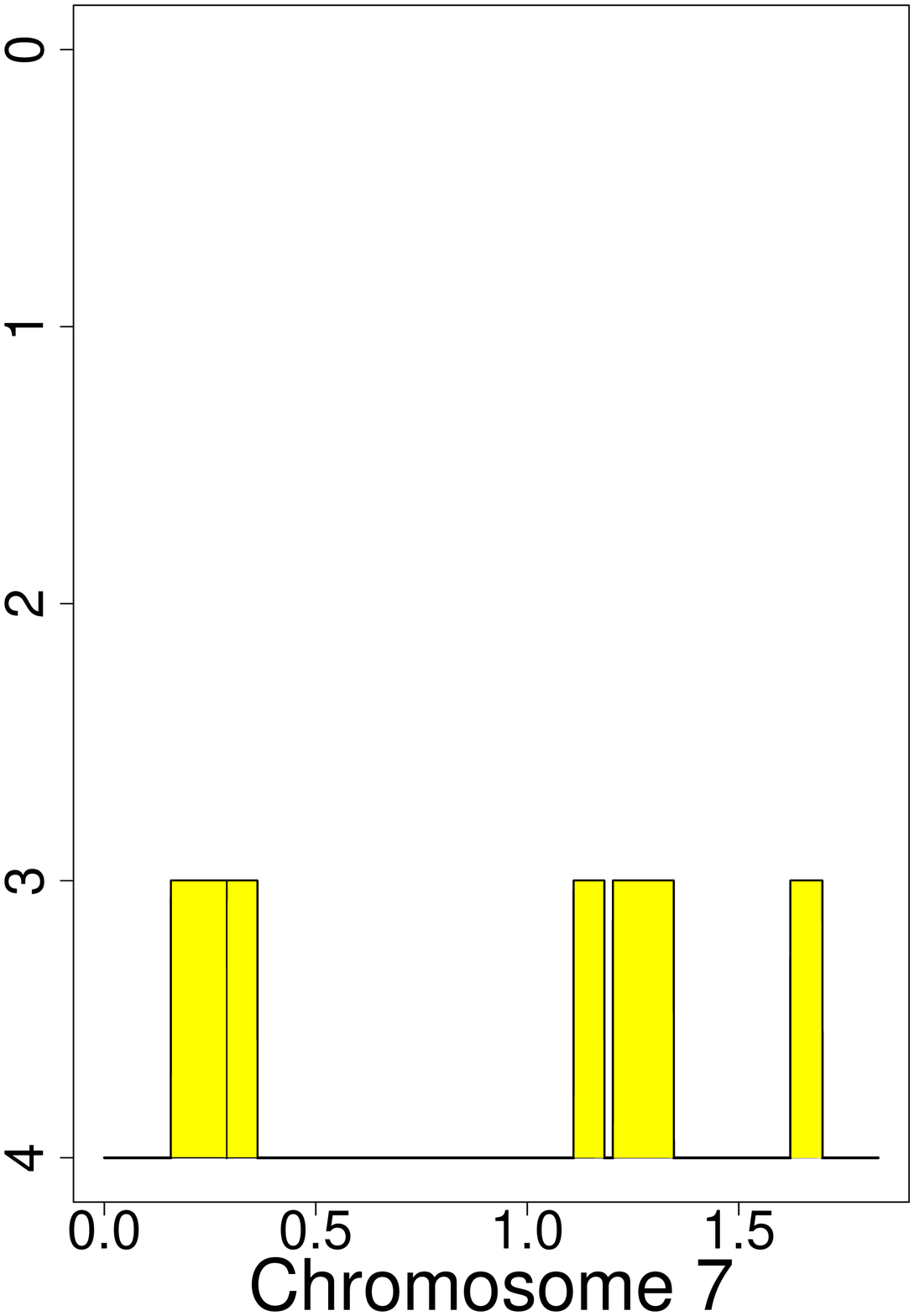}
  \end{minipage}
 \hfill
 \begin{minipage}[t]{3cm}
 \includegraphics[width=0.9 \textwidth]{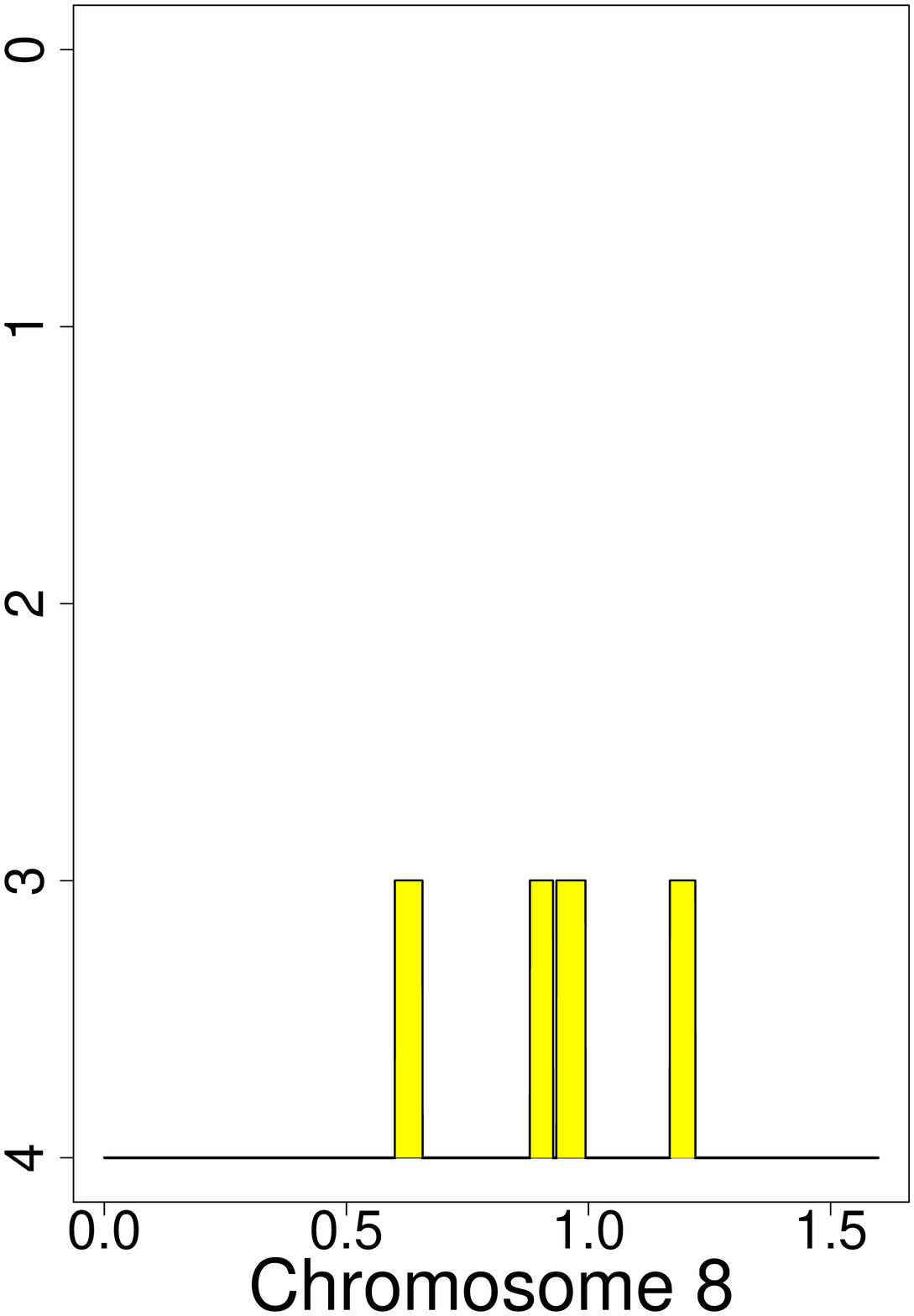}
 \end{minipage}
 \hfill
  \begin{minipage}[t]{3cm}
 \includegraphics[width=0.9\textwidth]{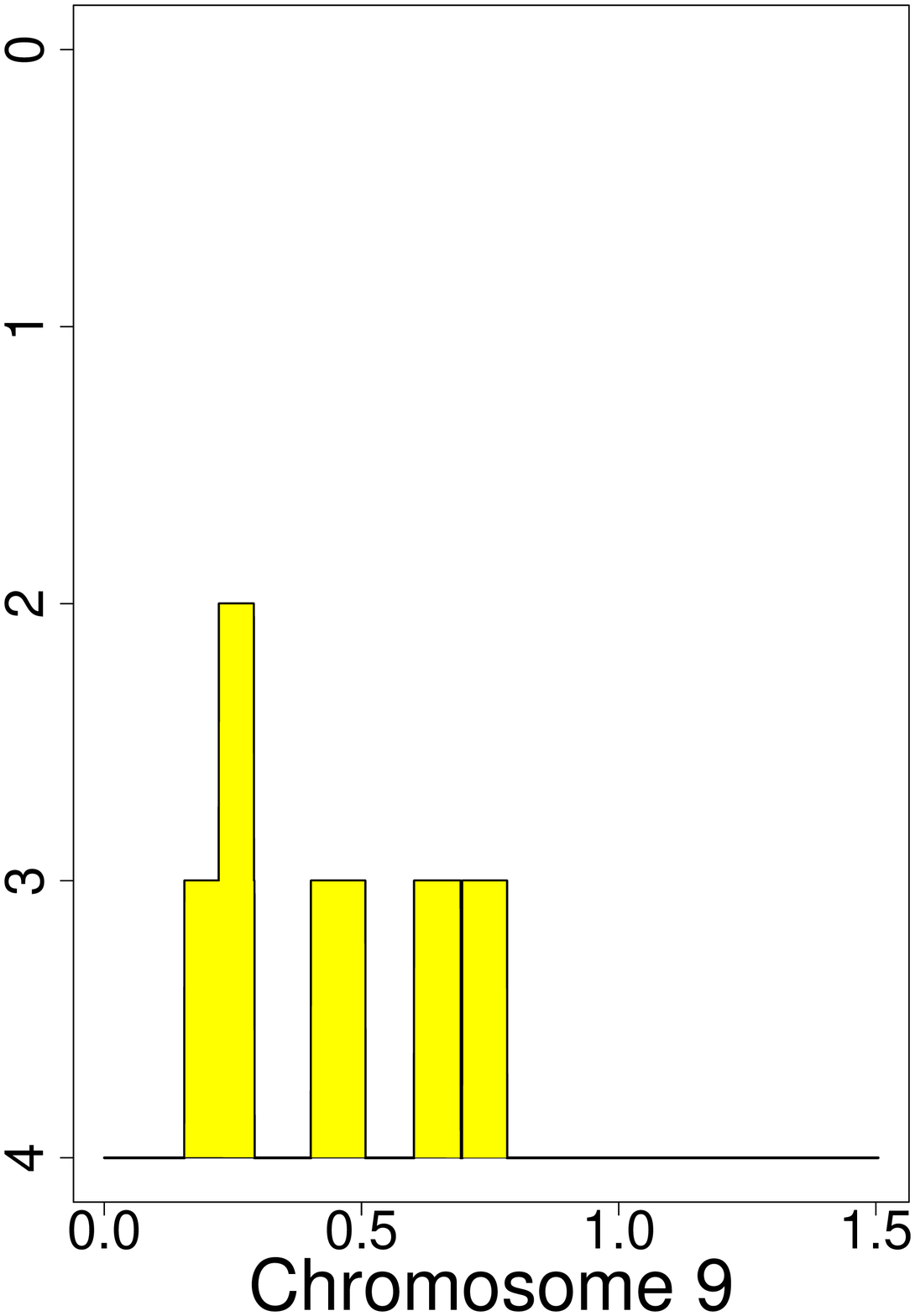} 
 \end{minipage}
 \hfill
 \begin{minipage}[t]{3cm}
 \includegraphics[width=0.9 \textwidth]{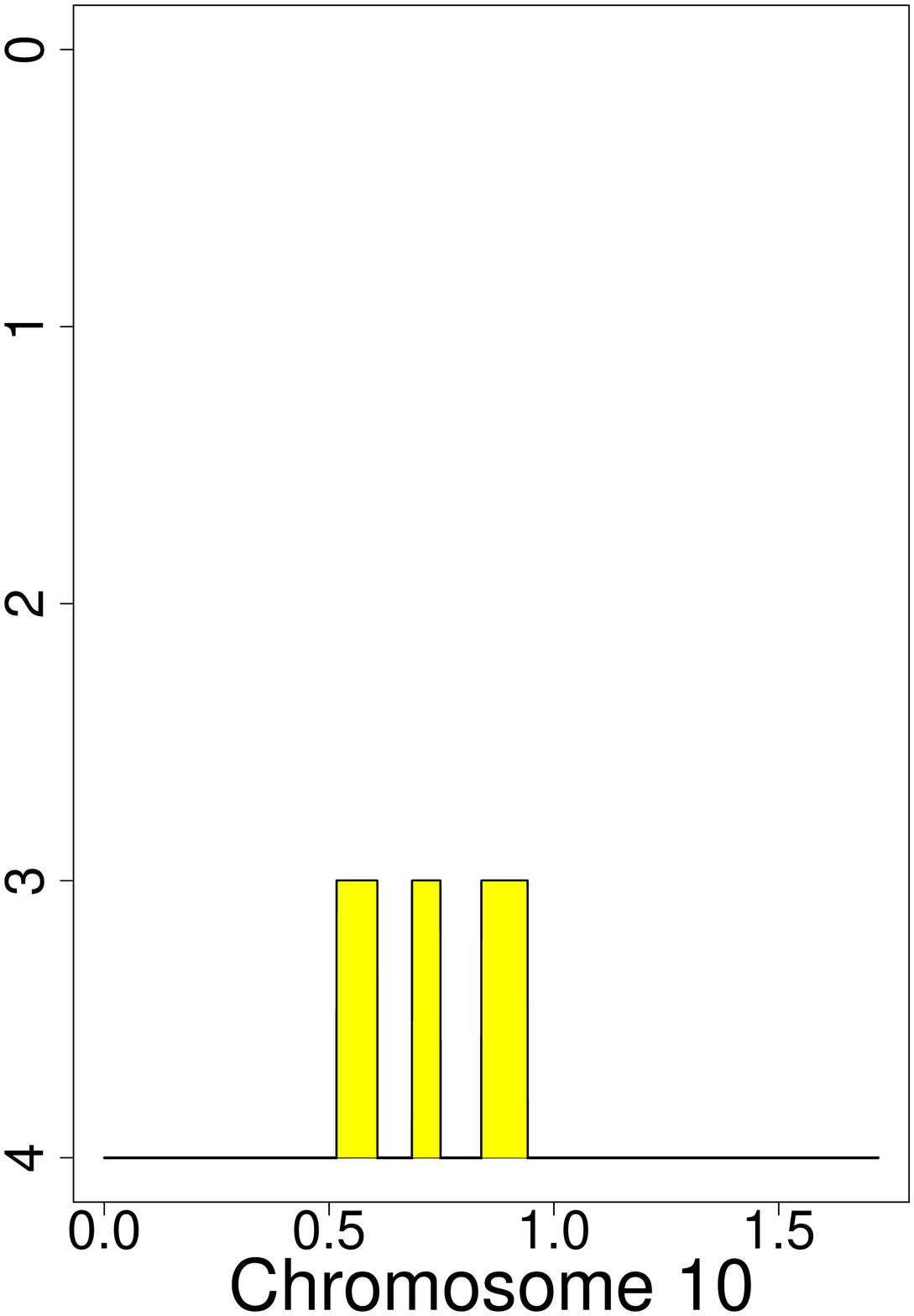}
 \end{minipage}
 \hfill
 \begin{minipage}[t]{3cm}
 \includegraphics[width=0.9 \textwidth]{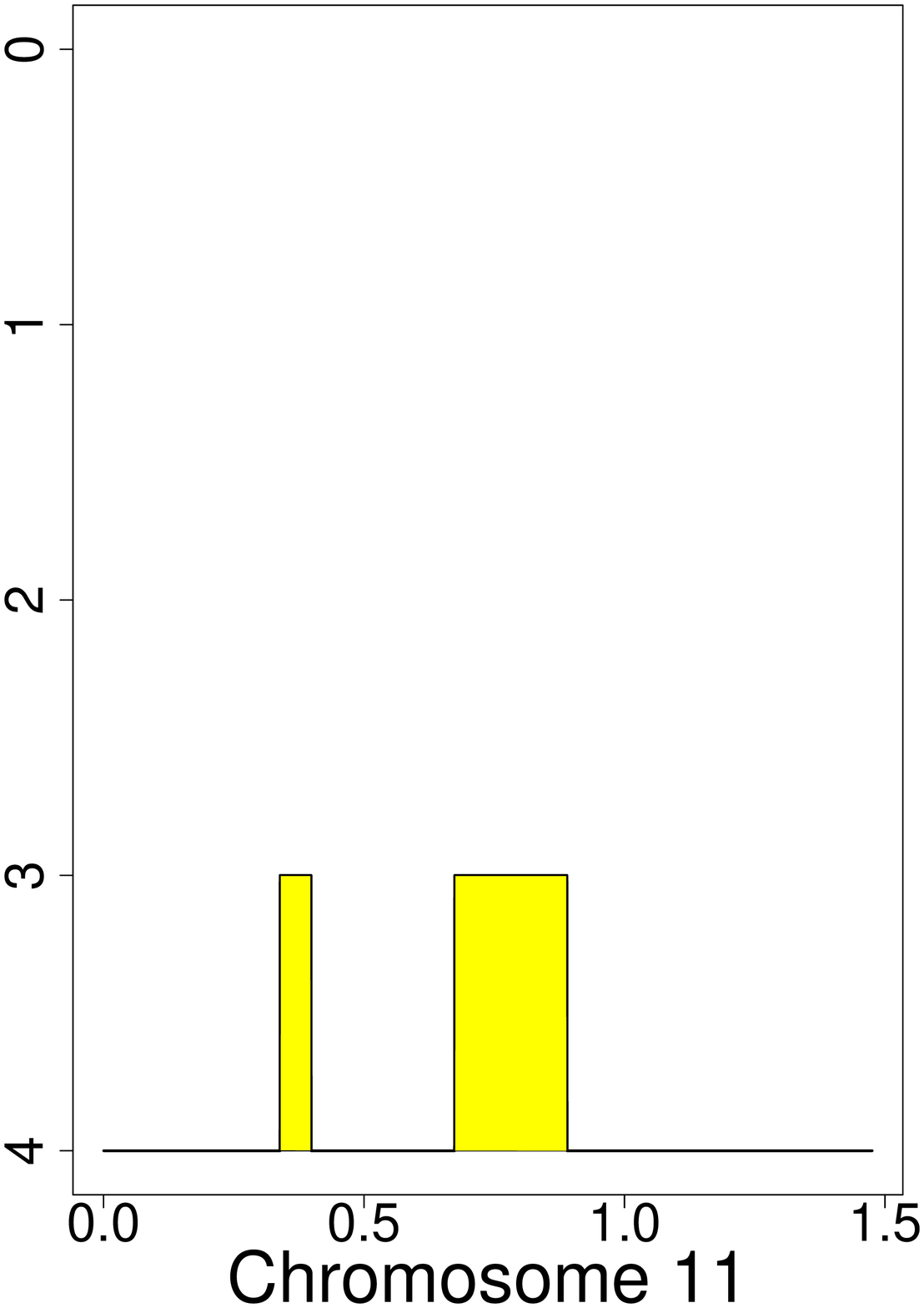}
  \end{minipage}
 \hfill
 \begin{minipage}[t]{3cm}
 \includegraphics[width=0.9 \textwidth]{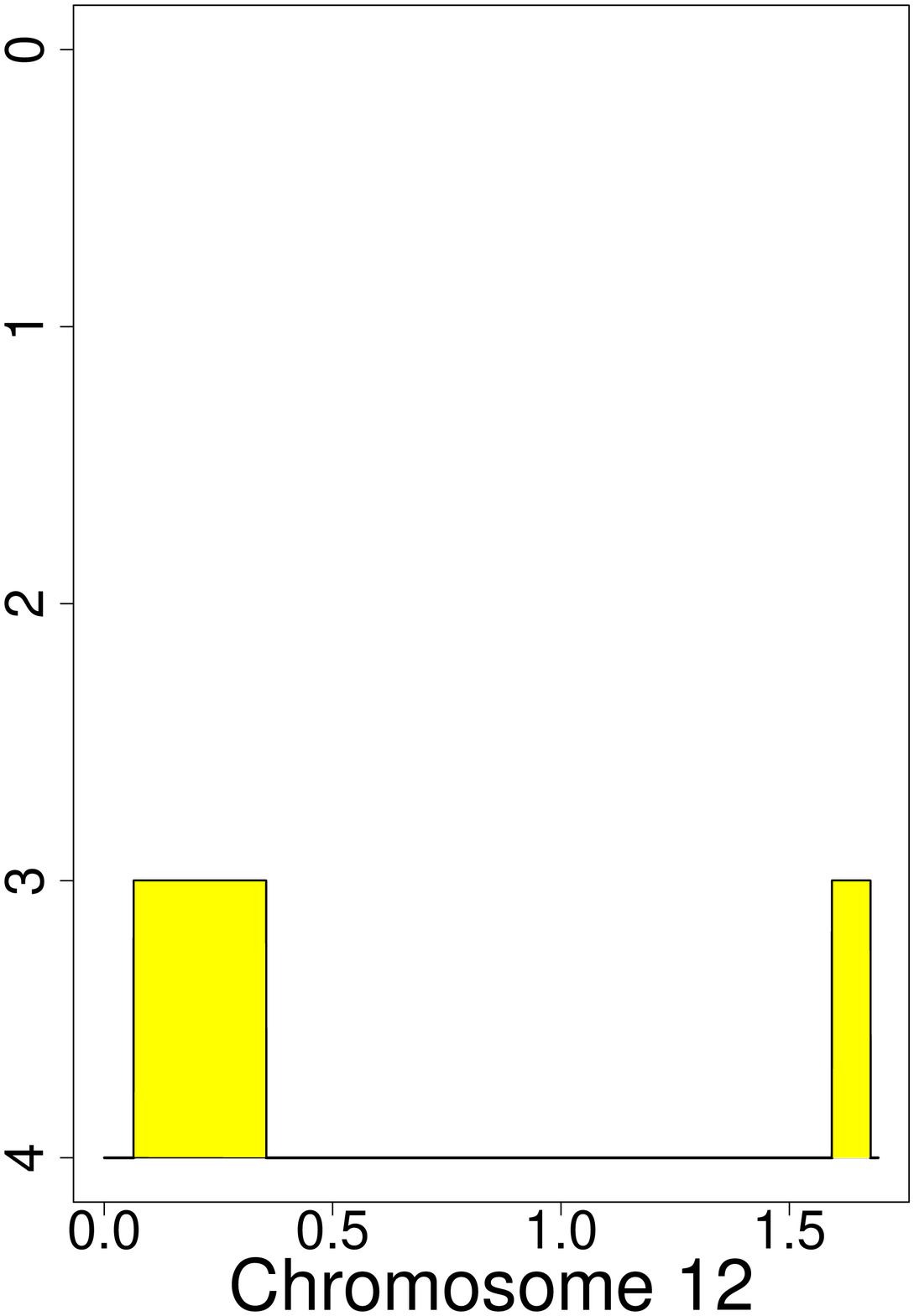}
 \end{minipage}
 \hfill
 \begin{minipage}[t]{3cm}
 \includegraphics[width=0.9\textwidth]{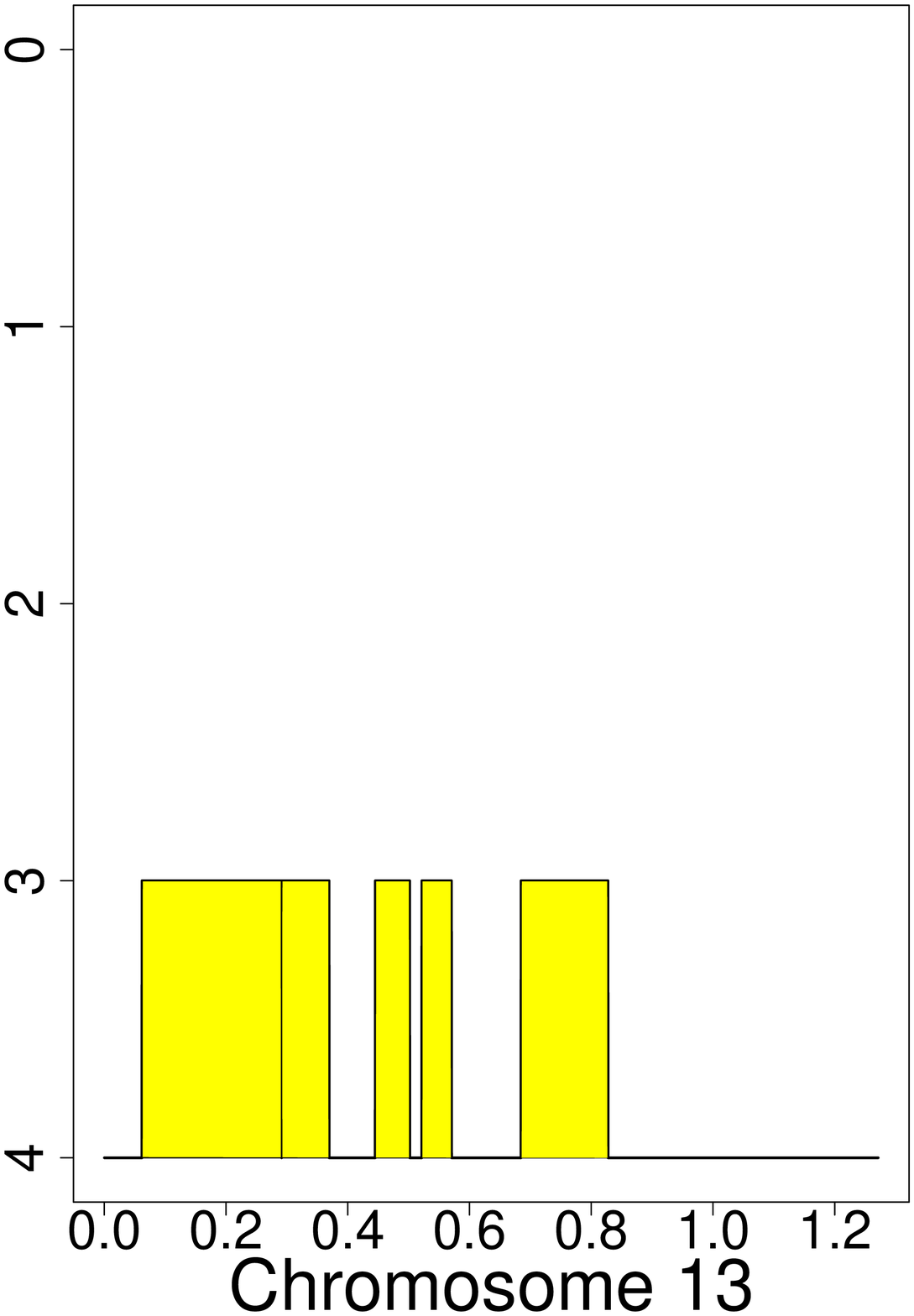} 
 \end{minipage}
 \hfill
 \begin{minipage}[t]{3cm}
 \includegraphics[width=0.9 \textwidth]{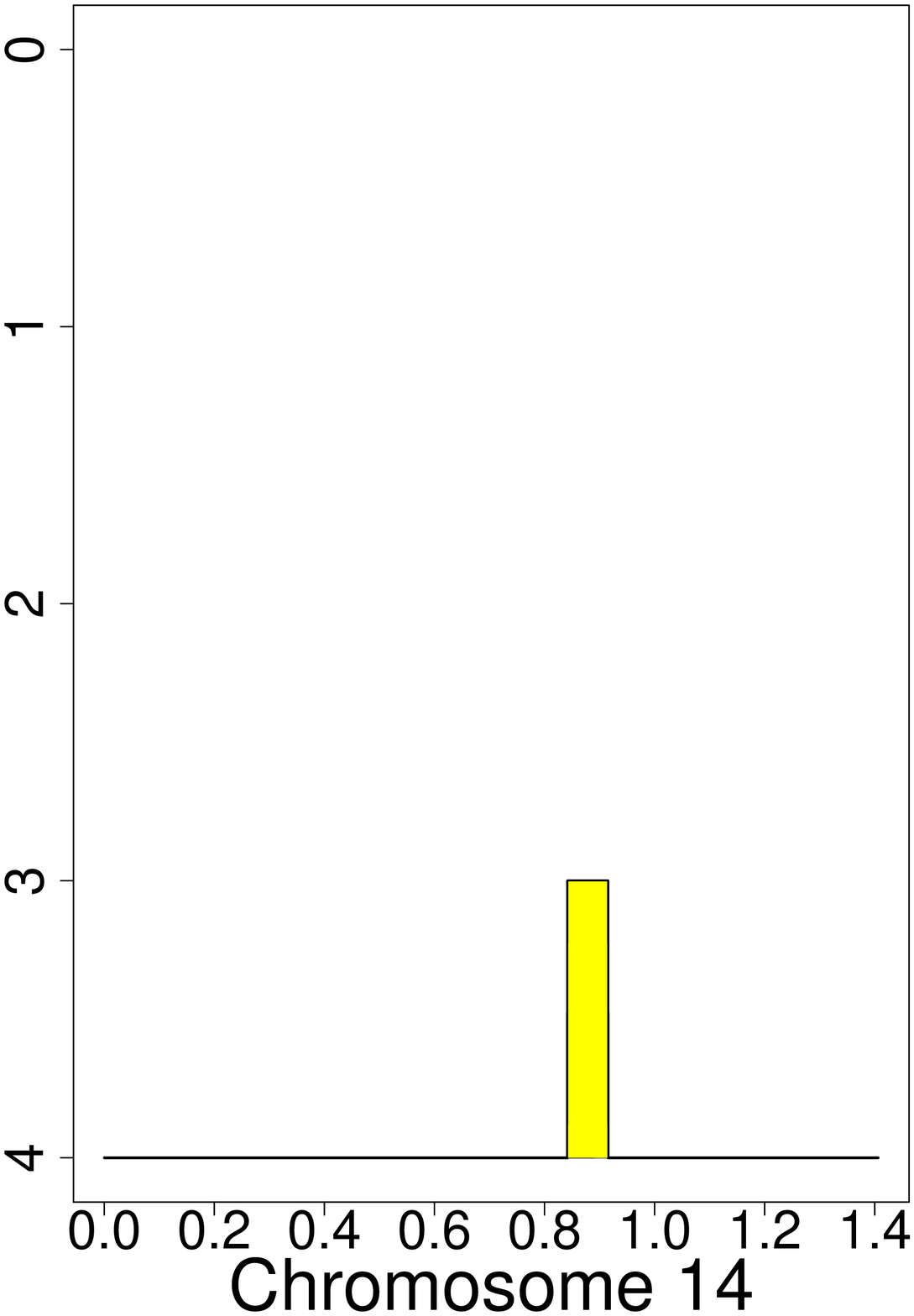}
 \end{minipage}
 \hfill
 \begin{minipage}[t]{3cm}
 \includegraphics[width=0.9 \textwidth]{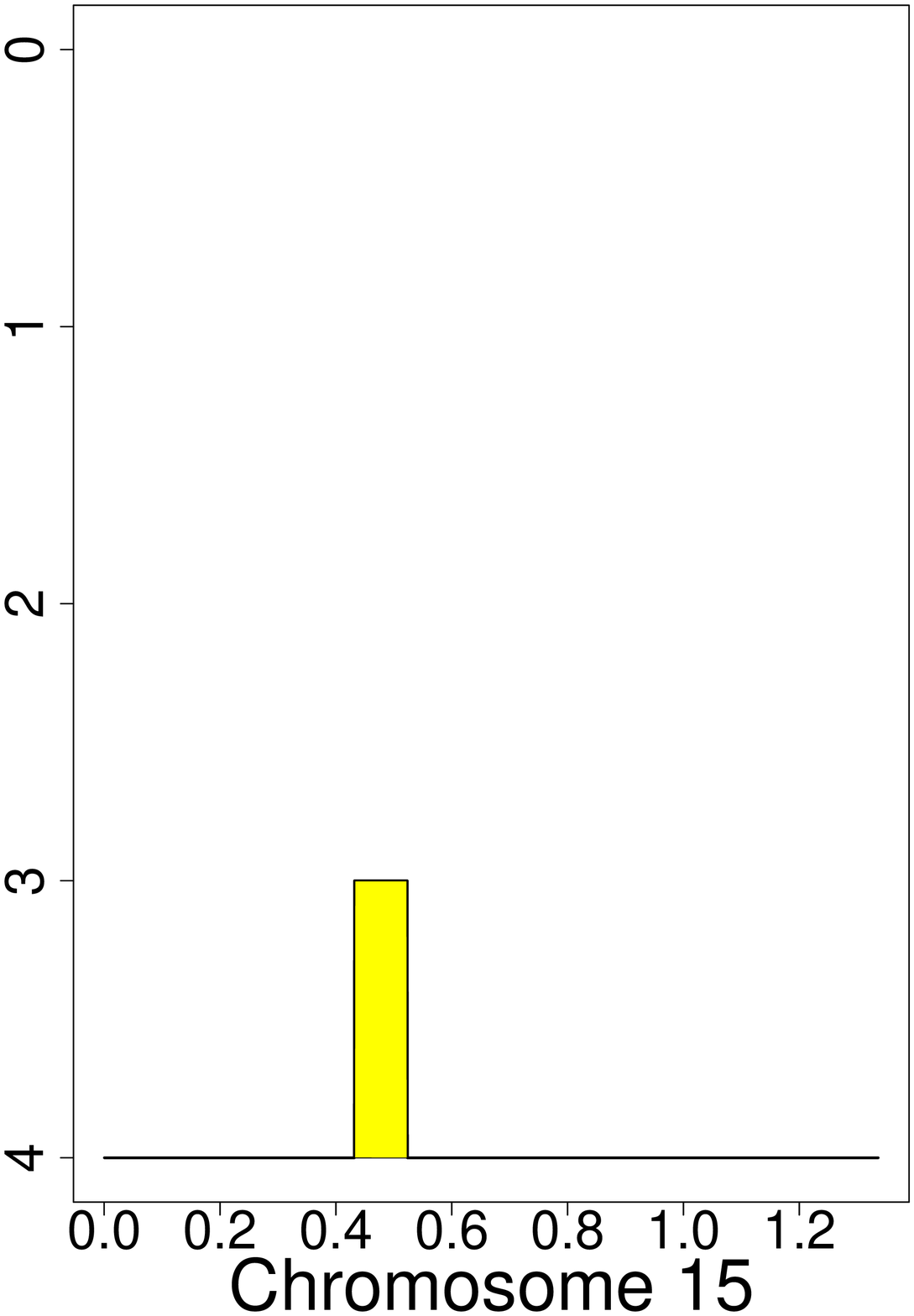}
  \end{minipage}
  \hfill
 \begin{minipage}[t]{3cm}
 \includegraphics[width=0.9 \textwidth]{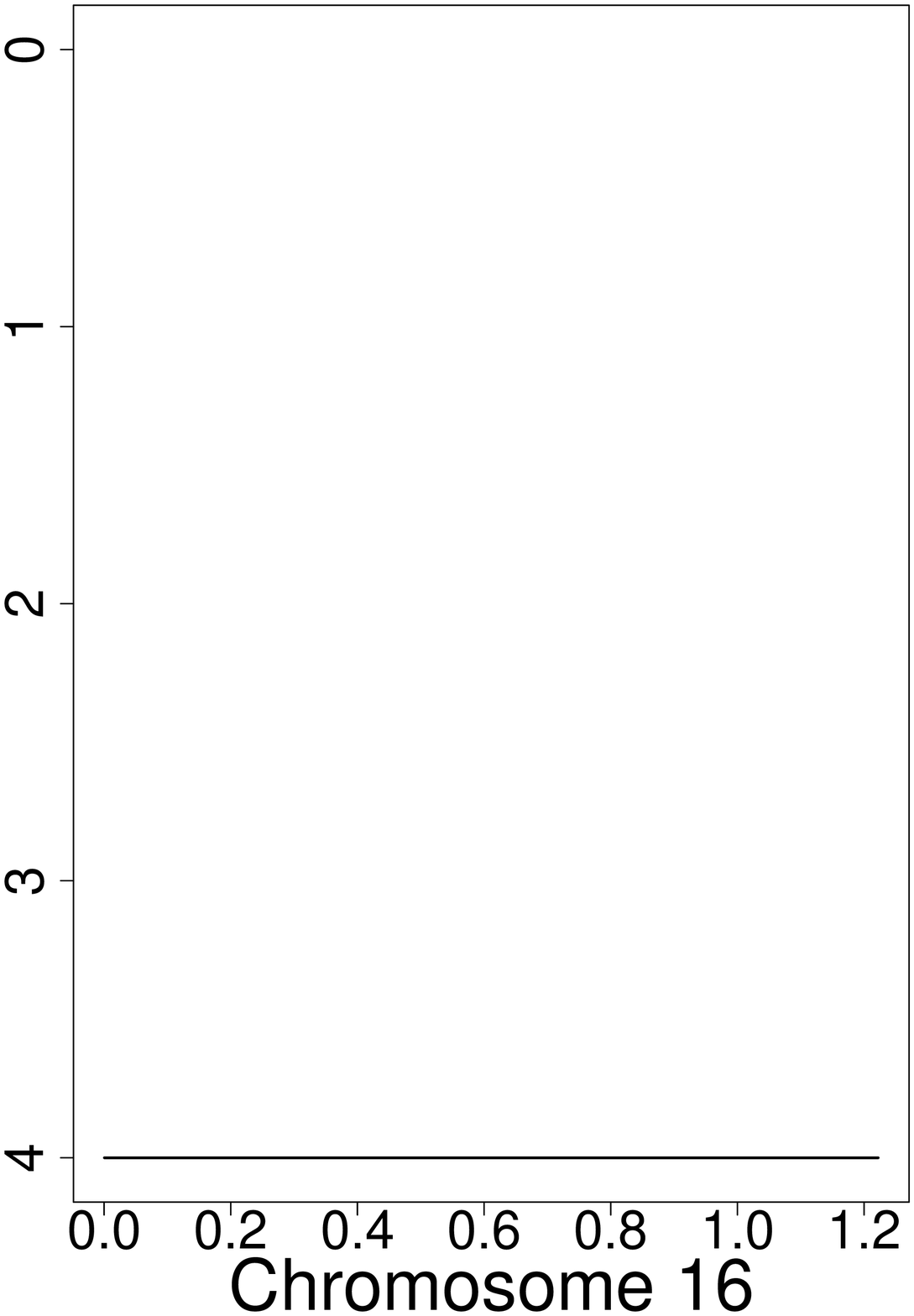}
  \end{minipage}
 \hfill
 \begin{minipage}[t]{3cm}
 \includegraphics[width=0.9 \textwidth]{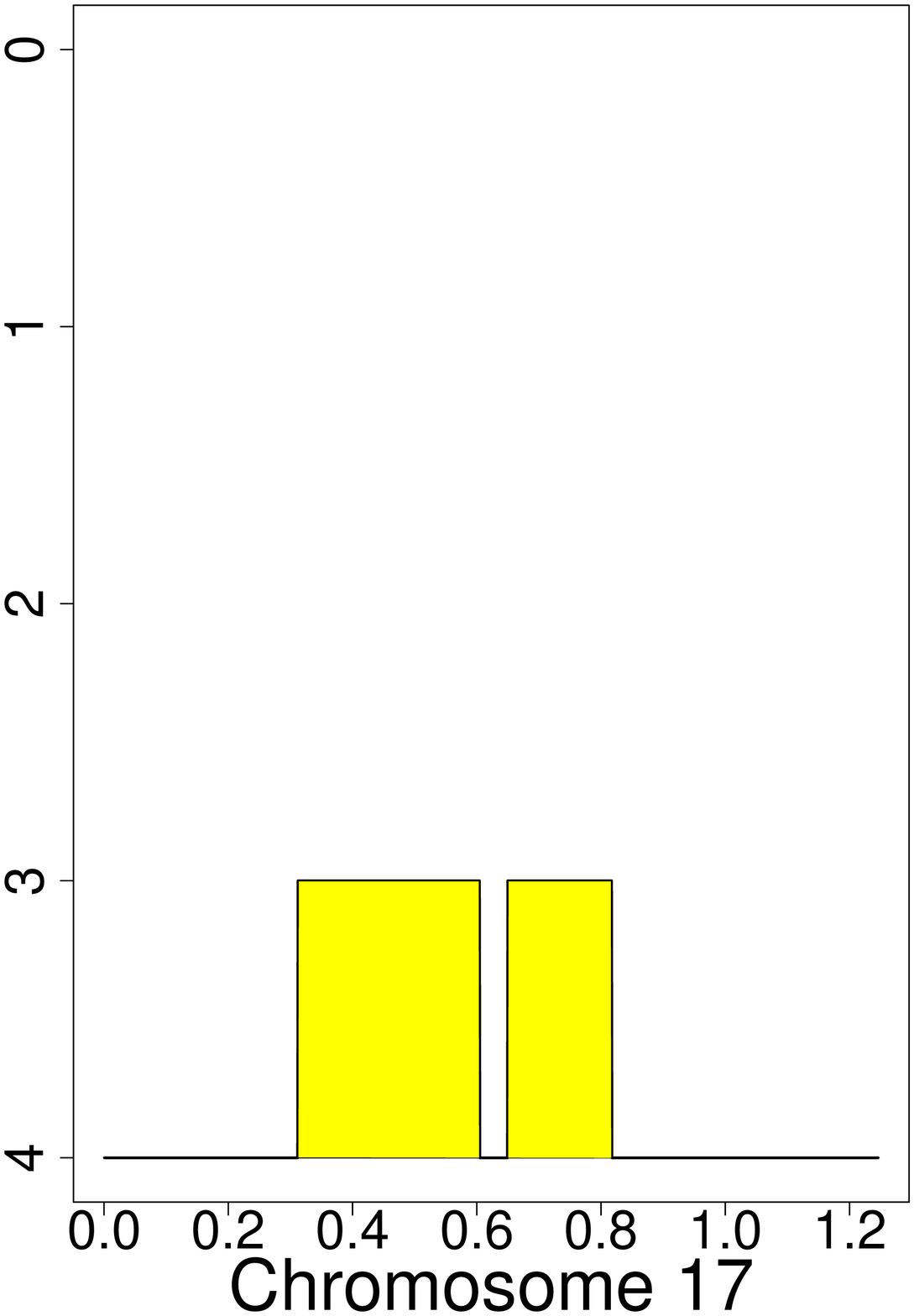}
 \end{minipage}
 \hfill
 \begin{minipage}[t]{3cm}
 \includegraphics[width=0.9\textwidth]{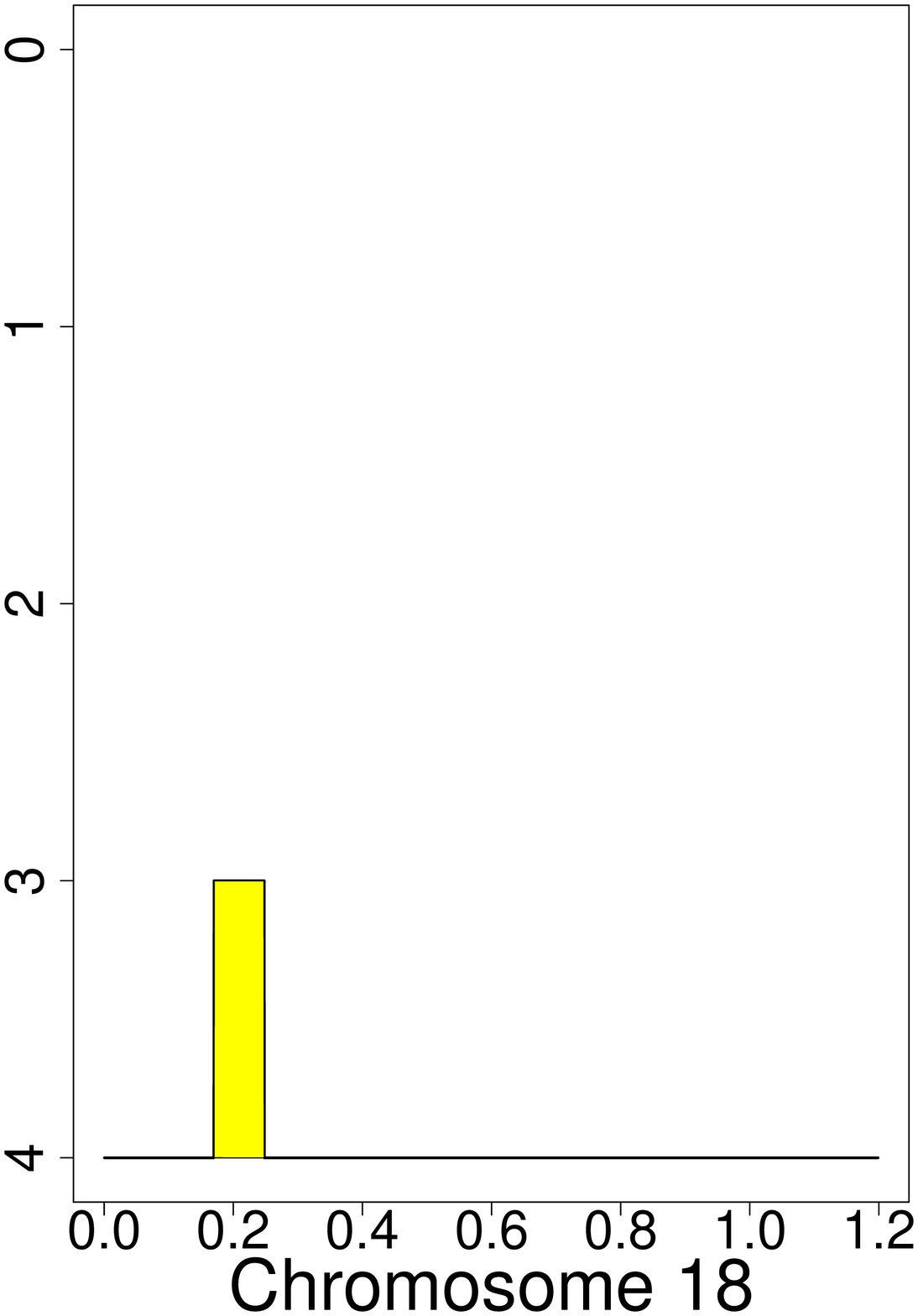} 
 \end{minipage}
 \hfill
 \begin{minipage}[t]{3cm}
 \includegraphics[width=0.9 \textwidth]{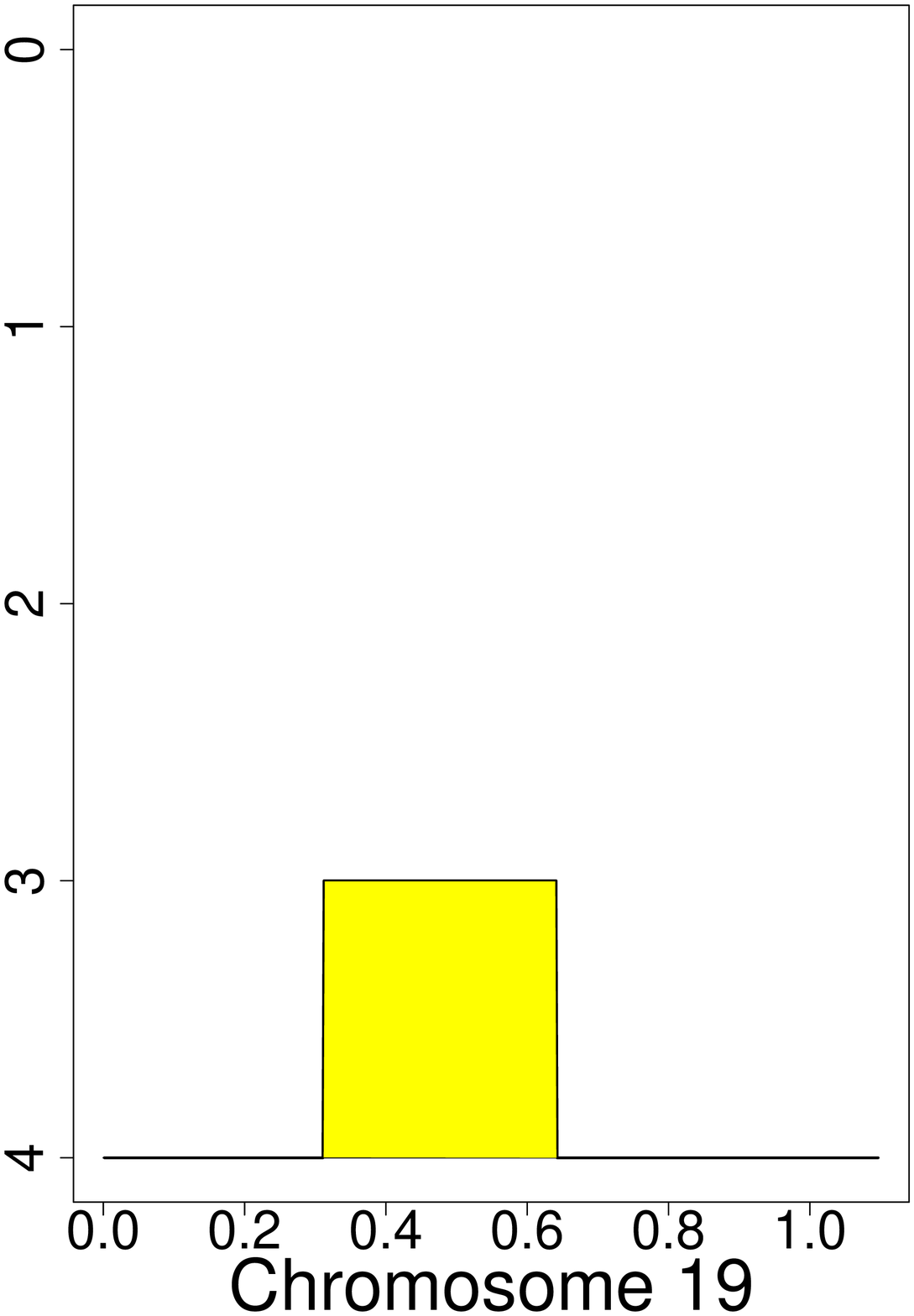}
 \end{minipage}
 \hfill
 \begin{minipage}[t]{3cm}
 \includegraphics[width=0.9 \textwidth]{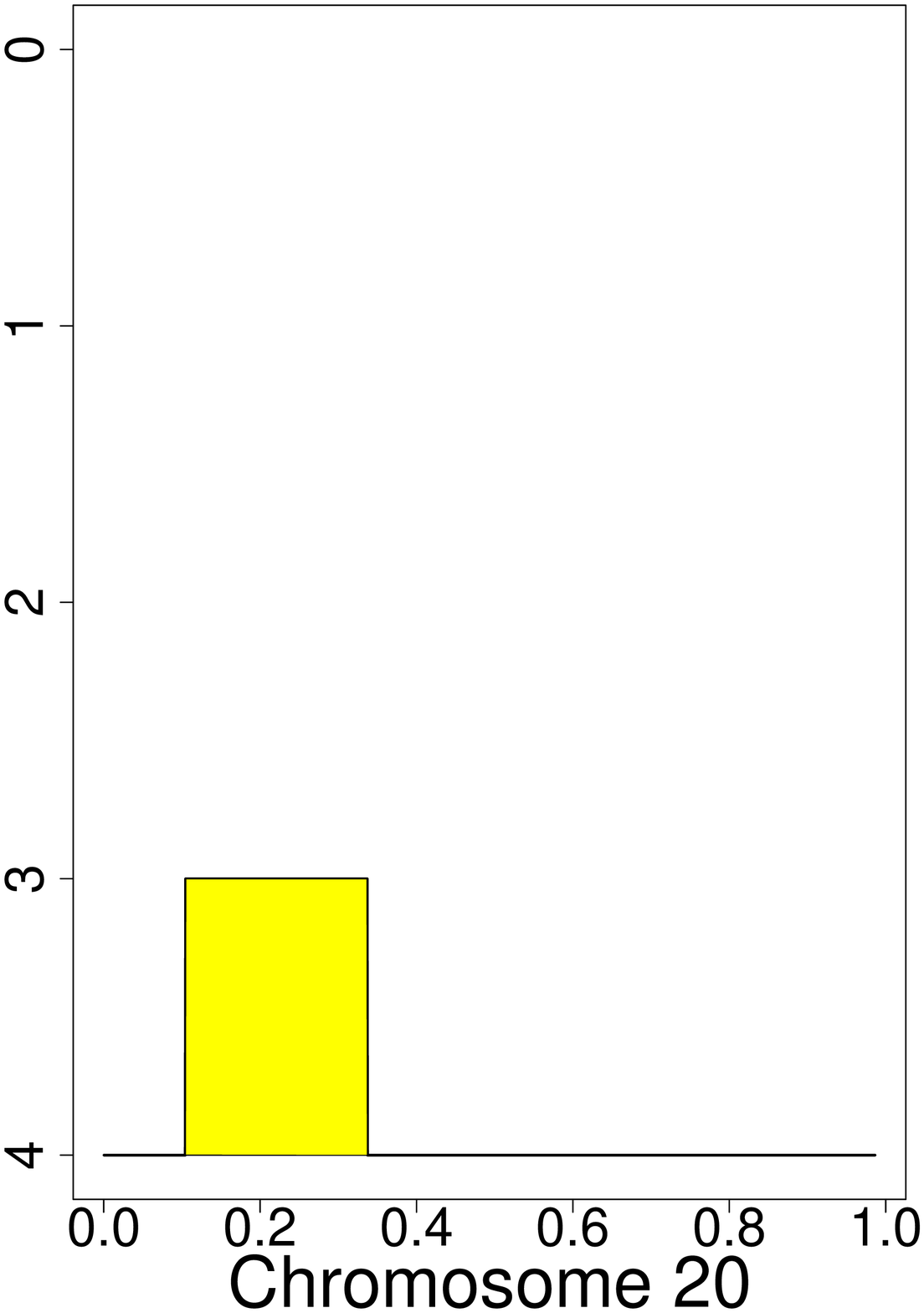}
  \end{minipage}
 \hfill
 \begin{minipage}[t]{3cm}
 \includegraphics[width=0.9 \textwidth]{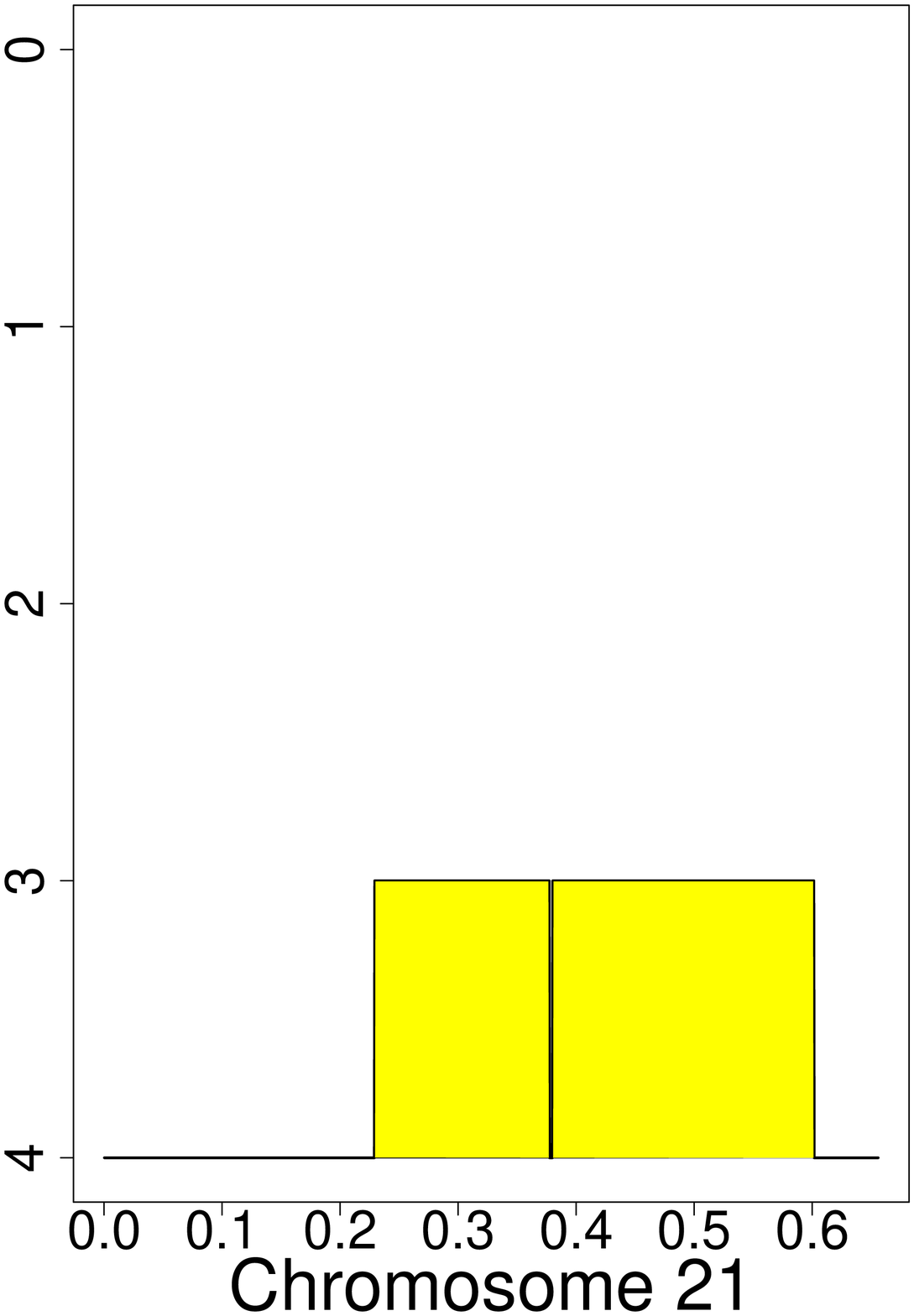}
 \end{minipage}
 \hfill
 \begin{minipage}[t]{3cm}
 \includegraphics[width=0.9 \textwidth]{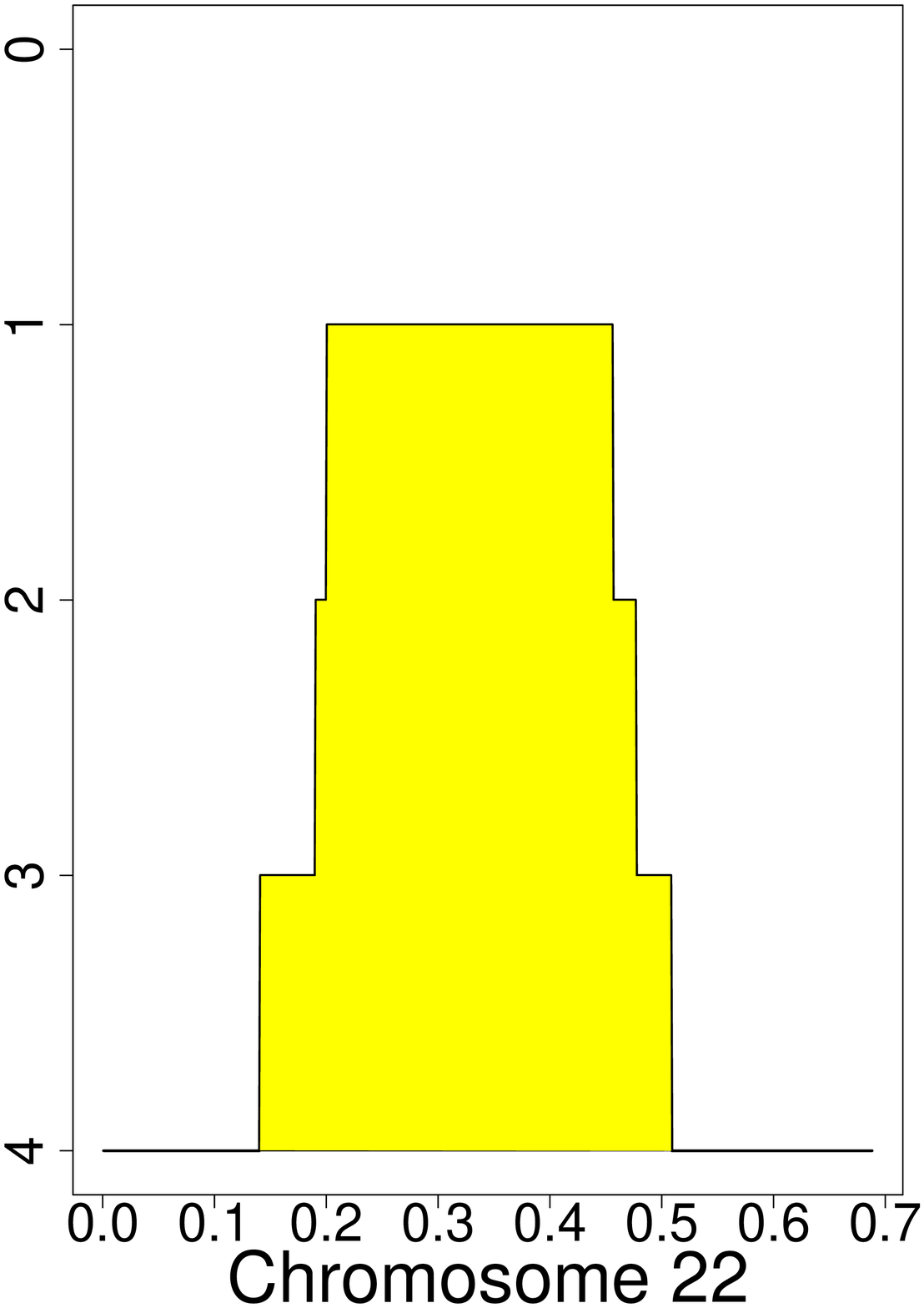}
 \end{minipage}
\caption{ \label{FMAll8}   
The black curve outlines $S_{200}(x)$  in black for the whole genome and the FG-FM family's actual SNP data.   
}
\end{figure}

Denser marker maps allow us to obtain  better and better approximations to  the true {Shadow}.  
  Figure~\ref{Increase}, shows  a sequence of such approximations for simulated 10k, 100k and 500k SNP data for  our FS-Z family.   We applied our 10k approximation to the real FS-Z 10k data, and found a unique interval on chromosome 11  where $S_{58}(x)$ took on its minimal  value of $0$ as seen in Figure~\ref{LodComp} (the choice of $58$ is discussed in section \ref{QVal1}). 
 As published in (\cite{Reiser}), this is the location of the  TRPC6 gene that harbors the disease causing allele.  The {Shadow} curve  $S_{200}(x)$  for the FG-FM family can be seen in Figure~\ref{FMAll8}.  Here we see a unique $S_{200}(x)=1$ interval on chromosome 22. 
In Section~\ref{PVal},  we see that  $P \approx \frac{1}{25}$ for such a region occurring somewhere in the genome,
 and so its existence is statistically significant and would be our best candidate for a disease harboring  gene locus.
 In the  Section~\ref{Cake1} we explore and sharpen this FG-FM candidate using the {Shadow Method}.

\subsection{Analysis of the {Shadow}}

 Notice in the definition of $S_M$ we only  consider partitions which are consistent with a streak of length greater than  or equal to $M$.     We encounter two potential problems when choosing $M$.  For $M$ too large, we run the risk of false negatives.  We quantify this with what we call {\it $Q$-Value} as  introduced in Section~\ref{QVal1}.
For $M$ too small, we encounter false positives,  as  discussed in Section~\ref{Noise1}.
In Section~\ref{Estimates} we see that using these notions we can make sensible choices for $M$.  In Section~\ref{Estimates}  we will also see that as the number of SNPs gets larger it will be possible to choose $M$ so that there is simultaneously a very small chance of a false positive and a very small chance of a false negative.

 \subsubsection{ False Negatives  \label{QVal1}}
Figure~\ref{Increase} shows that for the 10k and 100k SNP sets there are regions where $S$ exaggerates how far $x$ is  from being in  an IBD region, a situation that will lead to {\em false negatives} in our hunt for disease loci.
 In fact, in both the 10k and 100k examples we see that the method entirely missed the small IBD region to the left of the IBD region harboring the disease-causing allele at $x=1$. 
We would like to compute the  probability that we miss the true disease-allele harboring region.  We call this probability the  $Q$-Value and  find   in Section~\ref{QVal} that if we define
   \[  G = \sum ch_i \]
   \[ N = \#(SNP markers) \] 
then   we have 
 \[ Q = 1- \left(1+ \frac{M B G}{N} \right) e^{- \frac{M B G  }{N} }. \]
  Fixing the  $Q$-Value is a very natural way to choose $M$.  For example, in Figure~\ref{Increase} for the 100k simulation we chose $M = 103$ since this corresponds to $Q=0.05$ and for 500k we chose $M=217$ since this corresponds to $Q=0.01$ .   We chose $M=58$ for the 10k data since it corresponds to detecting a region that is at least as long as the expected length of a disease causing region.  
  
  A reasonable question might be: Why did we not simply choose them all so that $Q=0.01$?  The problem is that then the 10k and 100k analyses will  then  become cluttered with false positives, the subject of the next section.

 \subsubsection{  False Positives  \label{Noise1}} 
 
\begin{figure}
\begin{minipage}[t]{4cm}
 \includegraphics[width=0.9\textwidth]{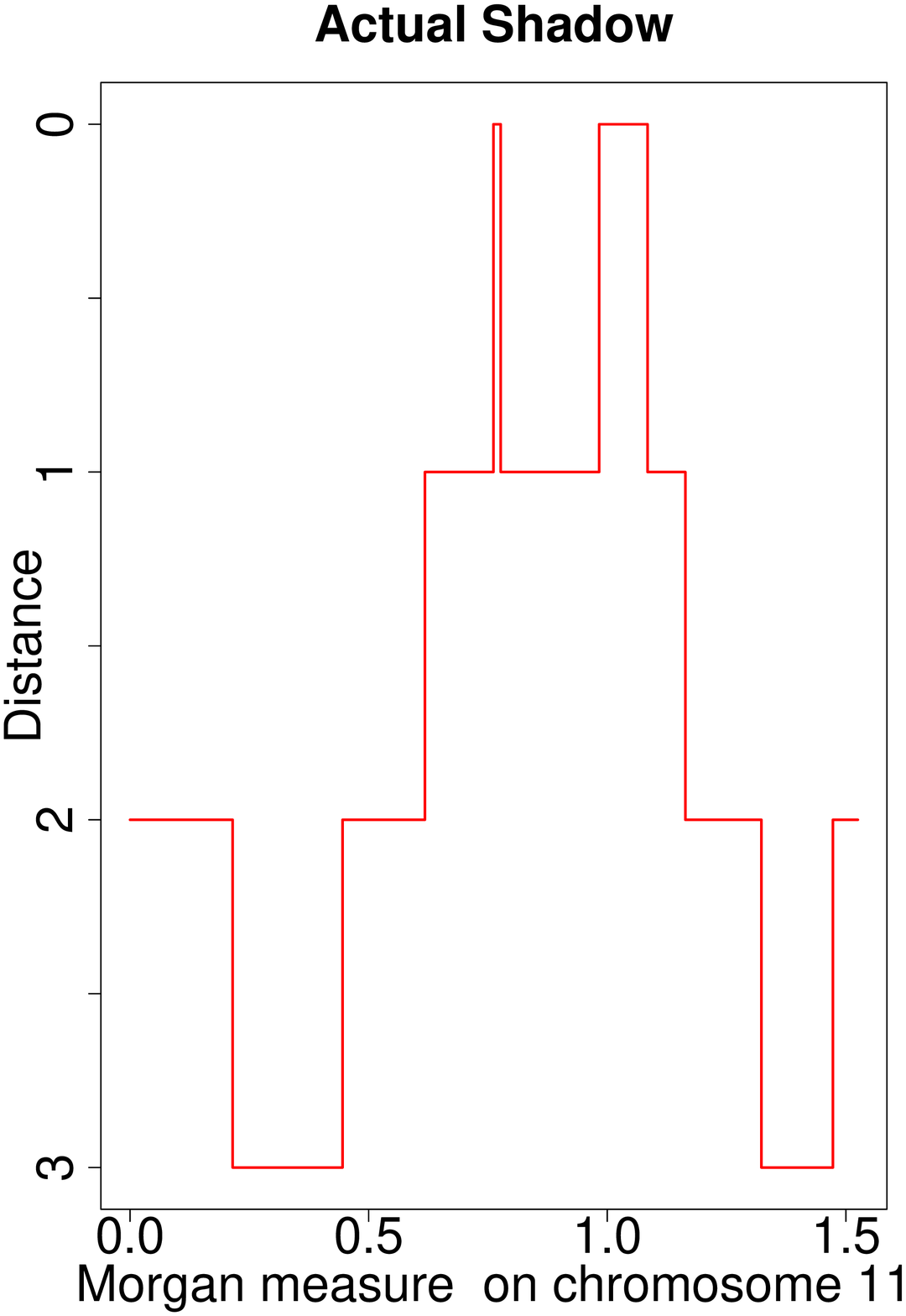} 
 \end{minipage}
 \hfill
 \begin{minipage}[t]{4cm}
 \includegraphics[width=0.9\textwidth]{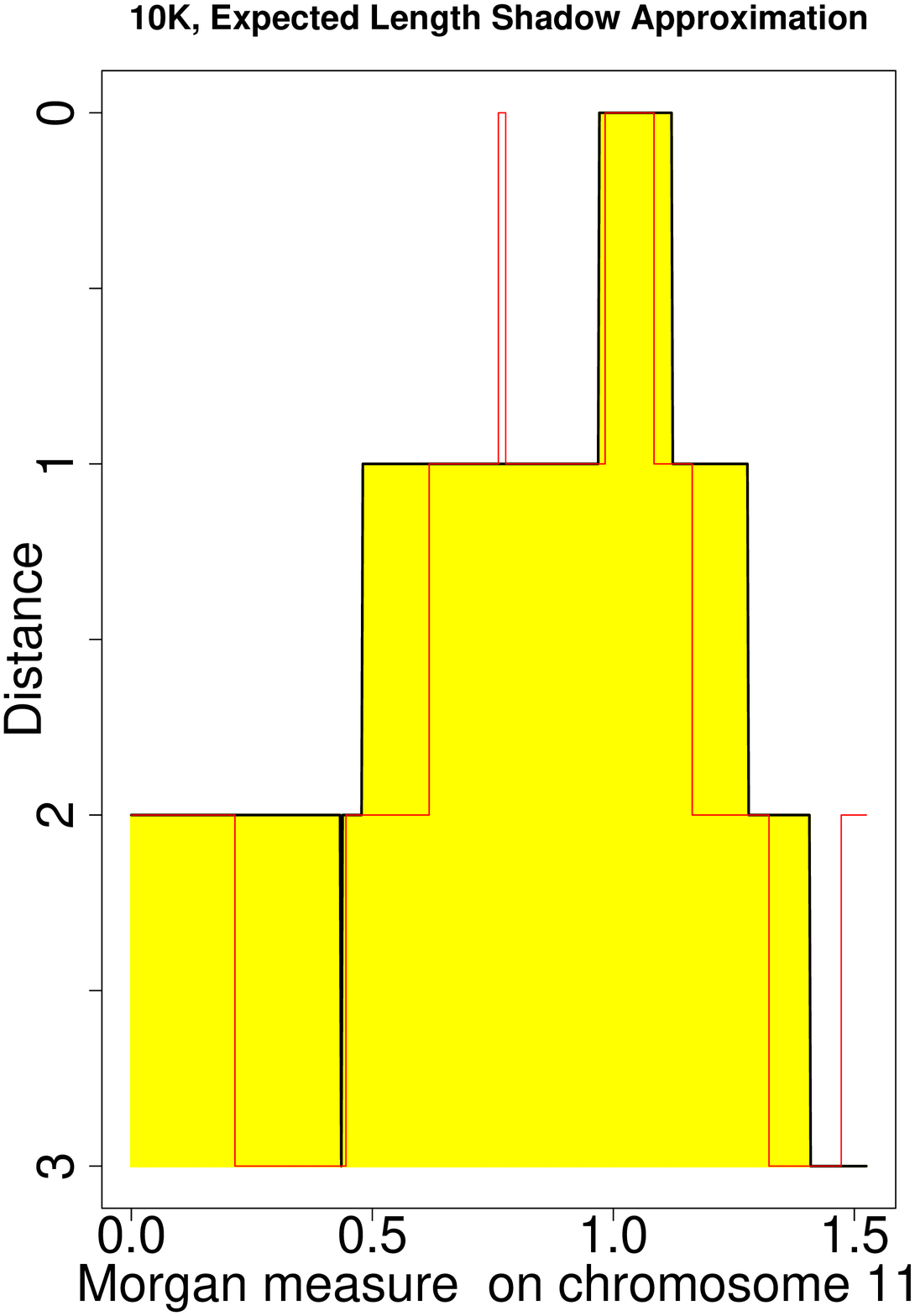} 
 \end{minipage}
 \hfill
 \begin{minipage}[t]{4cm}
 \includegraphics[width=0.9 \textwidth]{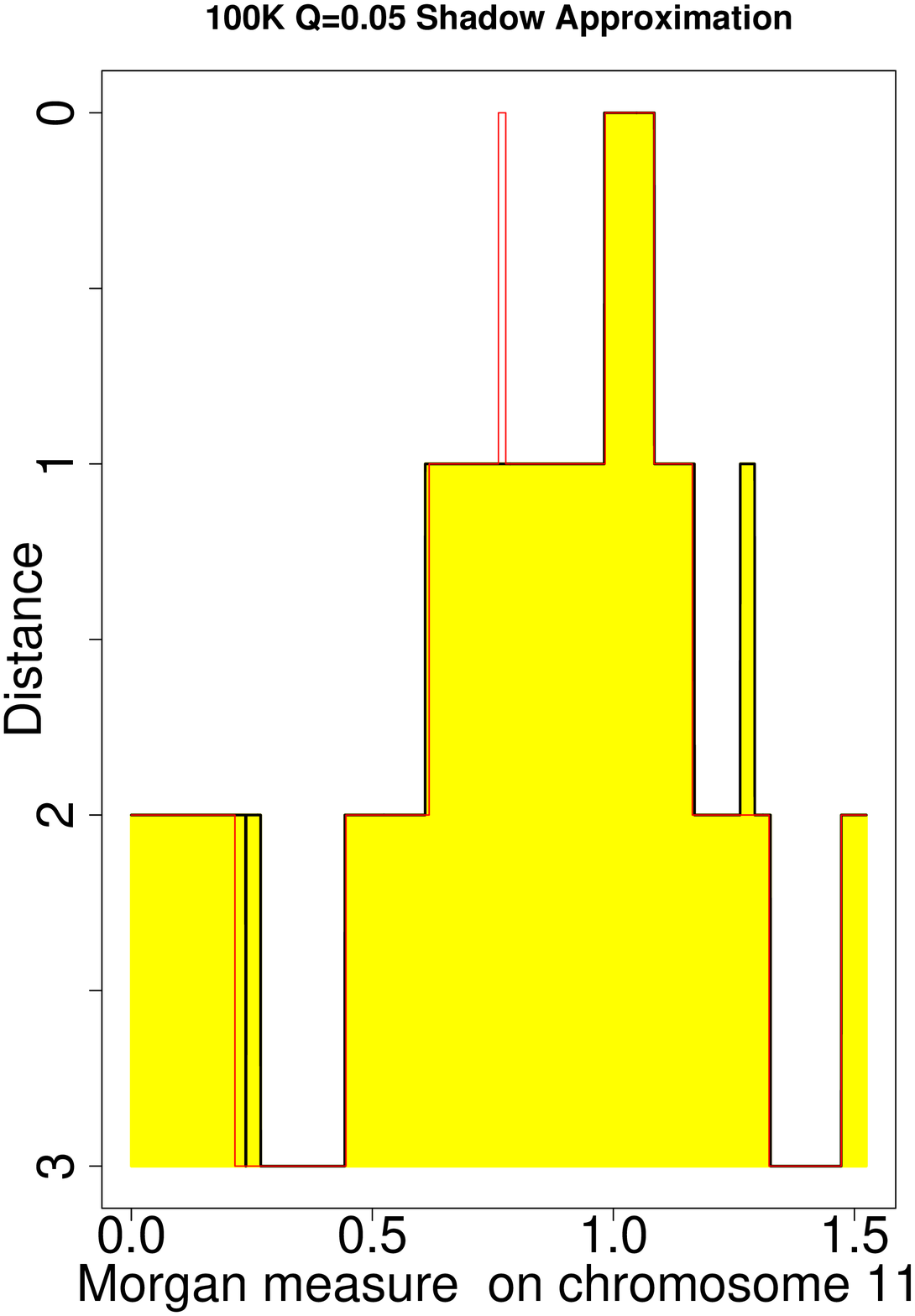}
 \end{minipage}
 \hfill
\caption{ \label{Increase}  
Here we see a simulation  the genetic process in the FS-Z family on chromosome 11 (see Section~\ref{PVal}).     We have plotted the simulation's  {Shadow}  using a red line. 
The black curves are approximations of the {Shadow} using: (A) 10k SNPs data, then (B) 100k SNPs data.
We choose our {Shadow} by simulating the genetic process 20 times and picked one with a   a second chance IBD region for illustrative purposes. 
  That we expect such chance IBD regions is  due in part to the fact that the $P$-Value for an IBD region is large for the FS-Z pedigree  (see Section~\ref{PVal}).  }
\end{figure}

\begin{figure}
\hfill
\begin{minipage}[t]{7cm}
 \includegraphics[width=.9 \textwidth]{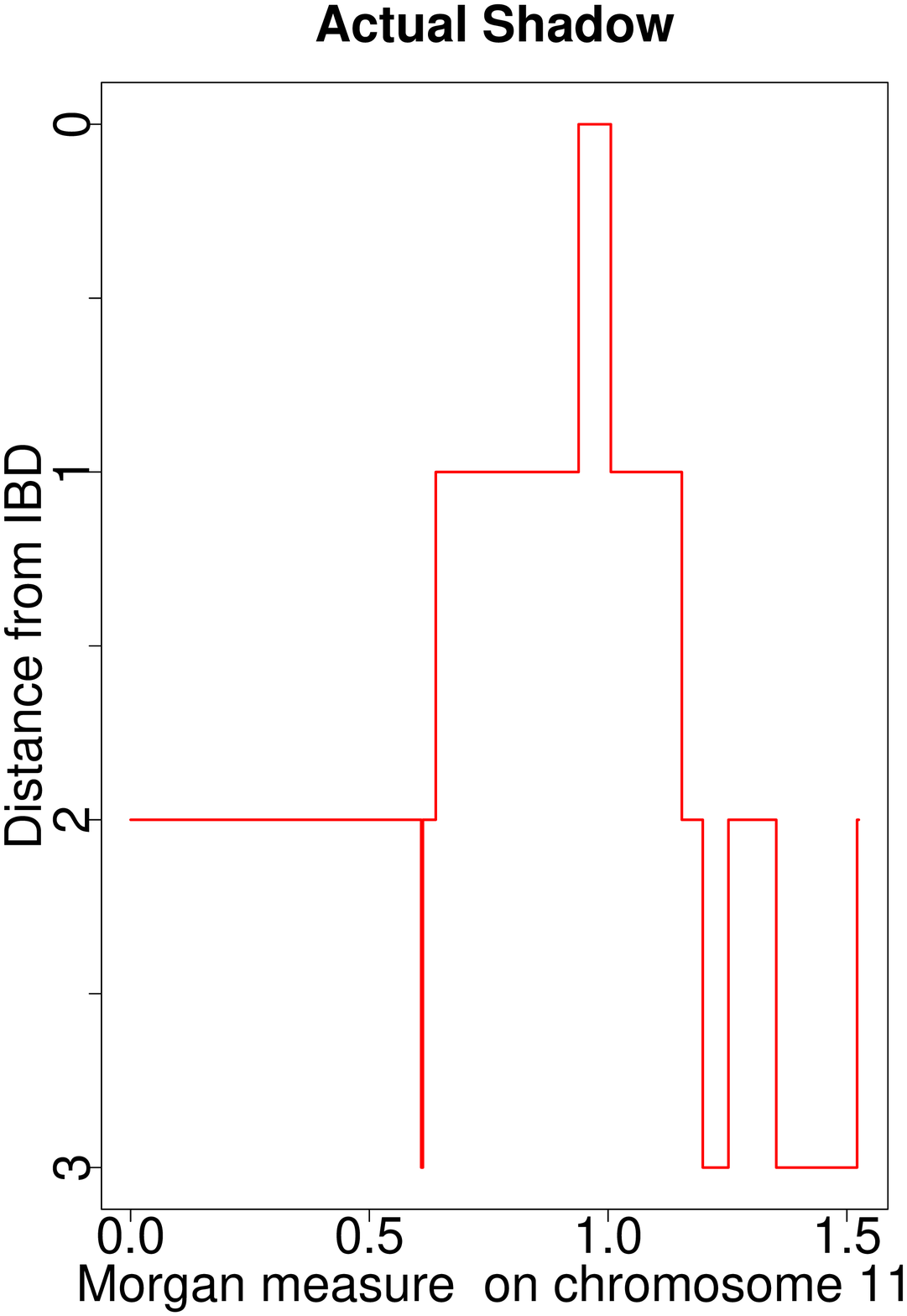} 
 \end{minipage}
 \hfill
 \begin{minipage}[t]{7cm}
 \includegraphics[width=.9\textwidth]{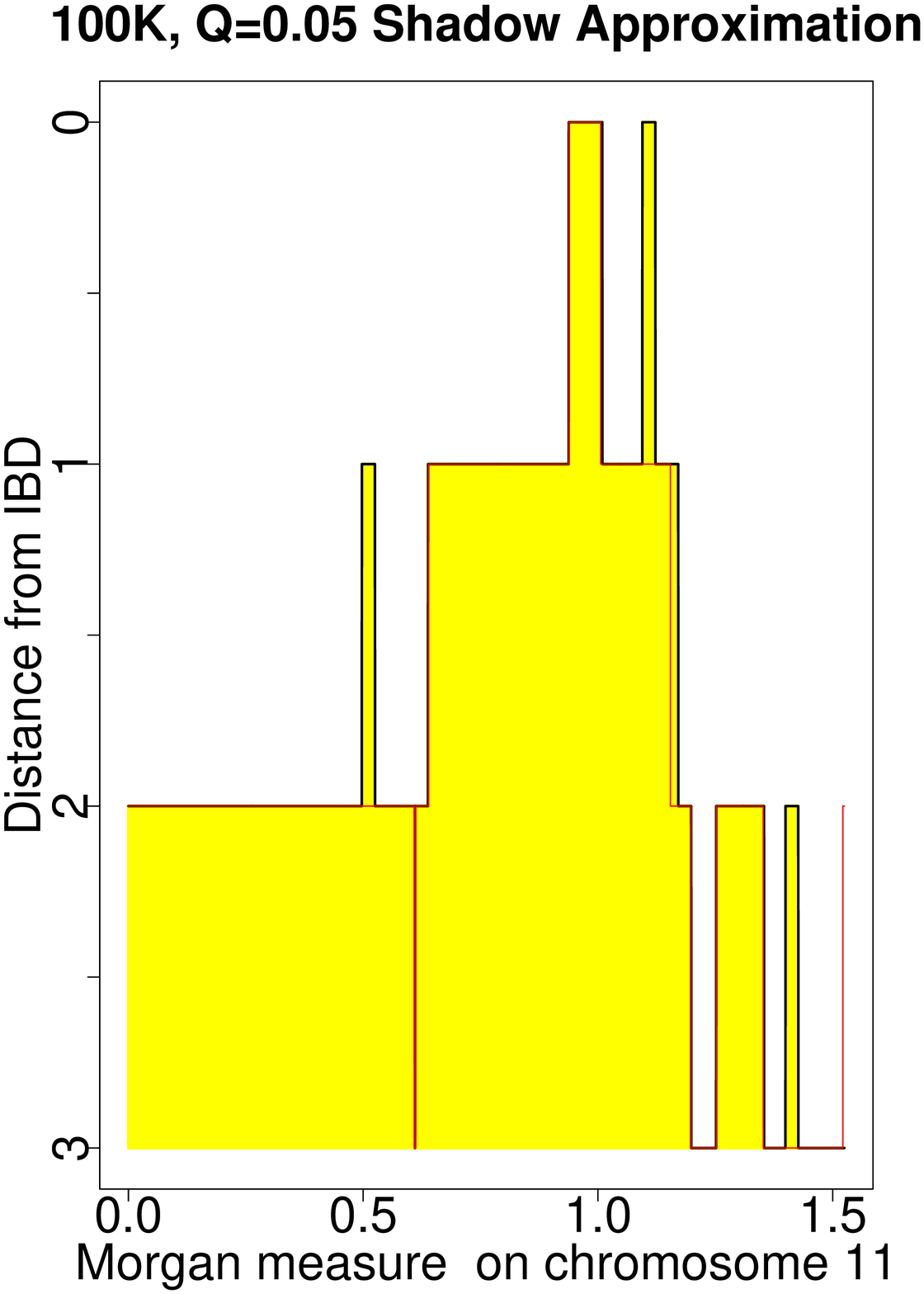} 
 \end{minipage}
 \hfill
 \caption{ \label{FalsePos} 
Here we see an example of a   false positive IBD region in our FS-Z family arising in a simulation. 
    }

 \end{figure}

Notice in Figure~\ref{Increase} that in the 10k and 100k cases there are regions where the value of $S$ exaggerates how close we are to being at an  IBD region, a situation that will lead to {\em false positives} in our hunt for disease loci. 
In Figure~\ref{FalsePos} 
we see an example of an $S_M=0$
 false positive region, and  we can estimate the 
probability of such a  false positive in the genome as follows.  
Let
\[ S= \#(\mbox{Samples for which we have SNP data}) \] 
 \[ p \approx P(\mbox{More Likely SNP  Allele}) \] 
\[ q=1-p \] 
 \[ p_j = p q ((1-p^{S-j})(1-q^{j})+(1-q^{S-j})(1-p^{j})) \] 
\[ p_{max} =  \max \{  p_j  \mid 1 \leq j  \leq \lfloor S/2 \rfloor  \}  \] 
\[ p_{min} =  \min \{  p_j  \mid 1 \leq j  \leq \lfloor S/2 \rfloor   \}  \] 
then we have that this   false positive rate $FP$ satisfies  
\[ FP <   N  p_{max} (1-p_{min})^M. \]

Notice, the $Q$-Value improves as $\frac{M}{N}$ decreases.  On the other hand, the larger the choice of  $M$  the smaller the false positive rate.  Hence there is a balance between making $M$ large in order to shrink the noise and  making  $\frac{M}{N}$ small in order to  shrink the $Q$-Value.    Explicit examples of this balancing act are given in Section~\ref{Estimates}.


  \onecolumn


\begin{figure}
\hfill
 \begin{minipage}[t]{7cm}
 \includegraphics[width=0.9 \textwidth]{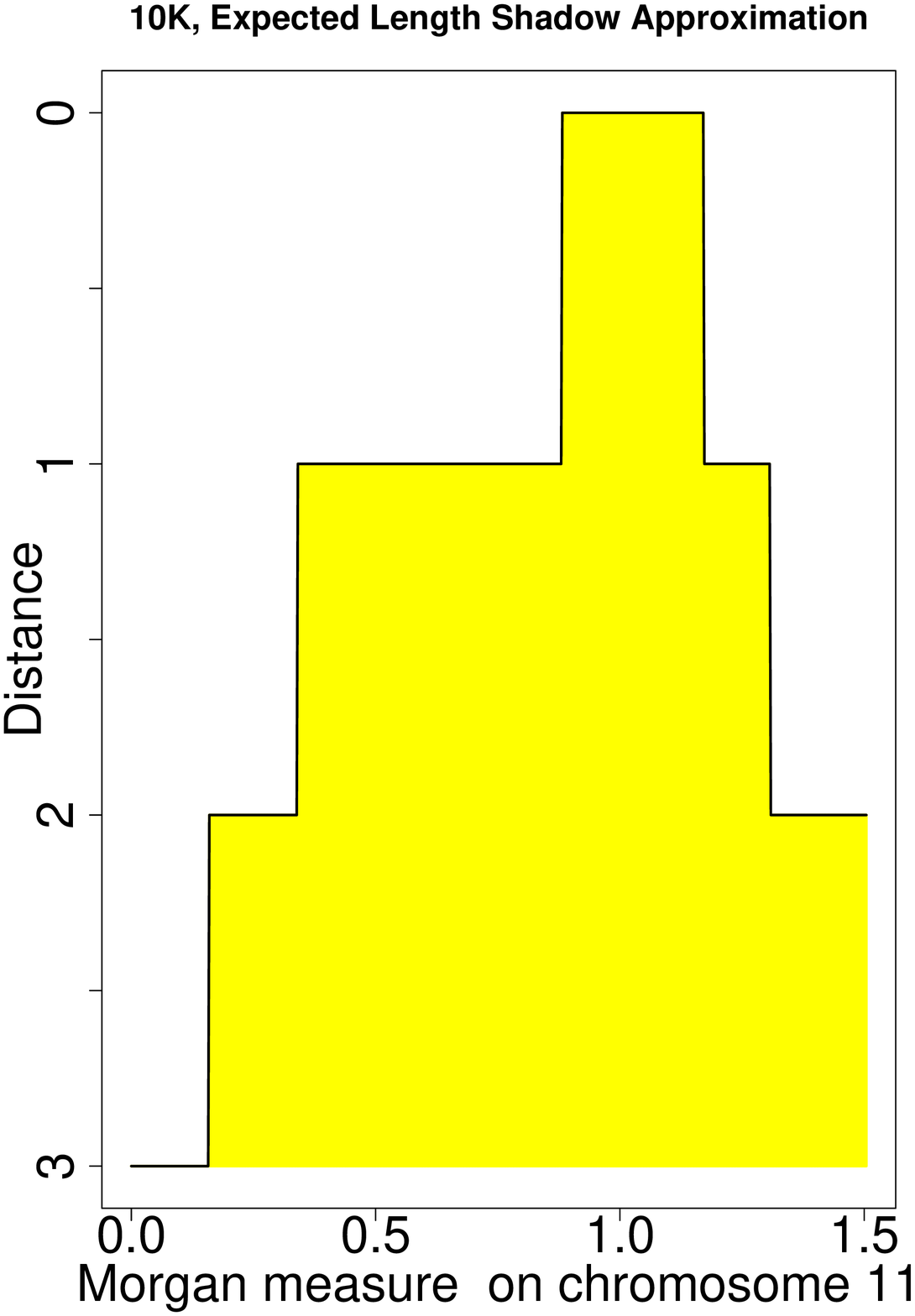}
 \end{minipage}
 \hfill
 \begin{minipage}[t]{7cm}
 \includegraphics[width=0.9 \textwidth]{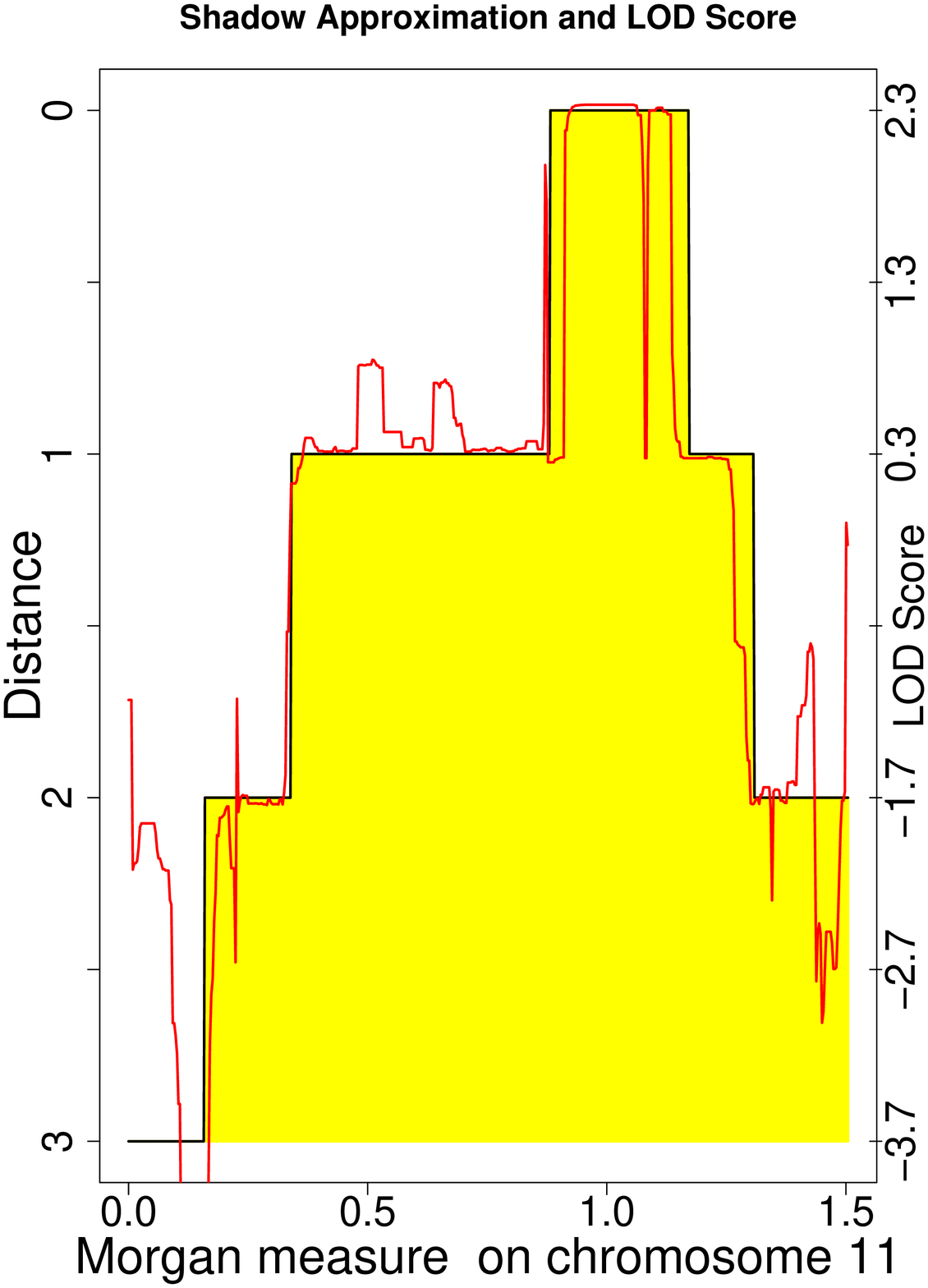}
 \end{minipage}
 \hfill
 \caption{ \label{LodComp} 
The black curve is   ${\it S}_{58}$ for the real  FS-Z family data  on chromosome 11.  In the next Figure, we compare  ${\it S}_{58}$ to   $LOD(x)$ in red as computed  by the Genehuneter2 program as implemented by the dCHIP program (see \citet{Kruglyak} and \citet{Leykin}).
 The LOD score values are on the right hand $y$-axis. 
    }
\end{figure}

\twocolumn

 \section{The {Shadow Method}  \label{Cake1}}

For both the 10k data on family FS-Z and the 100k data on the FG-FM family  it is impossible  to simultaneously make $Q$ and $FP$ small. By taking a more careful look at the data we are still able to
reduce the $Q$-Value.

  We use the fact that near a disease locus, $S$ has a very distinct tiered or ``wedding cake'' shape.   More precisely a true  $S=0$ region sits on top of an $S=1$ region  which sits on top of an $S=2$ region and so on, each layer requiring at least a pair of obligate crossovers to make the transitions between the tiers. In Figure~\ref{LodComp} we see a very typical example.       This structure allows us to detect an $S=0$ level by searching for a cake with a long  $S=1$ region as its top layer.  This technique will work best when such candidate regions are themselves rare, for example in the FG-FM family.      Armed with such candidates  we can take   a more detailed look at the definition of $S$.   Namely, we notice that when approximating the {Shadow}, at each point in the genome  we obtain a list of partitions of the samples that are  compatible with the data and  these partitions can be used to provide  greater insight into the disease loci.    It is the full use of this information that is called the  {\it Shadow Method}.   The analysis of these partitions takes two primary forms that we will now explore.

The first case is as in Figure~\ref{Flank}, where we see  an example of a  large $S=1$ region on chromosome 2 in the FS-Z data that looks a lot like a cake missing its top layer.   We find that the left half is given by the  partition consisting of $\{113,114,115\}$ and its complement, while the right half is determined by the singleton  $\{213\}$ and its complement, and in the middle the two partitions are both consistent.   This is indicated schematically on the right hand side of Figure~\ref{Flank}.  How can this happen?  The most likely possibility, as illustrated in Figure~\ref{Flank},  is that there is an IBD region separated by crossovers as indicated.  Whenever an $S=k$ region is comprised of a pair  partitions which differ by incompatible obligate crossovers  that intersects in a region compatible with the removal of these   obligate crossovers,  we can deduce the likely existence of an $S=k-1$ region.   

\begin{figure}
\hfill
 \begin{minipage}[t]{7cm}
 \includegraphics[width=0.75 \textwidth]{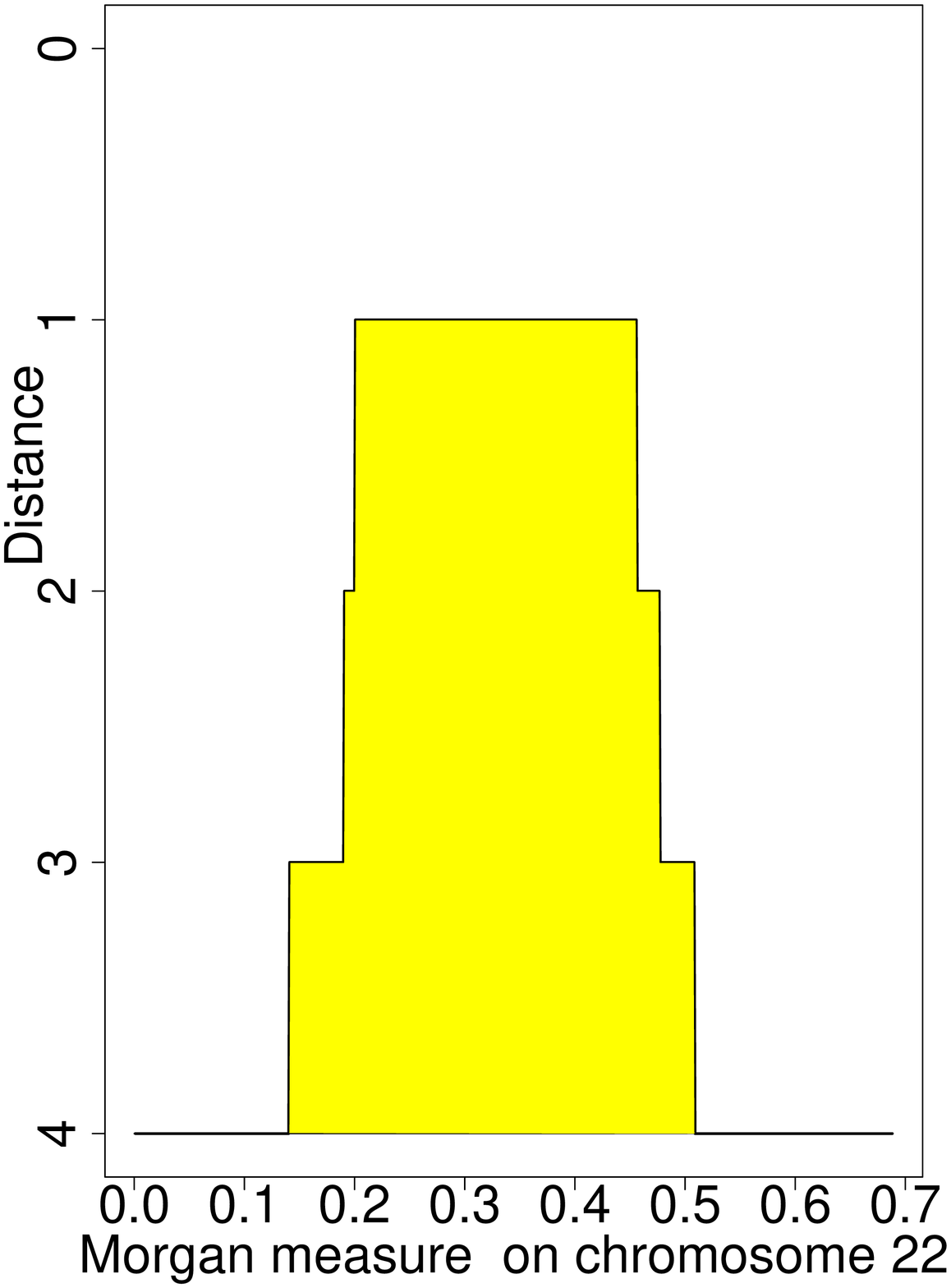}
 \end{minipage}
 \begin{minipage}[t]{9cm}
 \includegraphics[width=0.75 \textwidth]{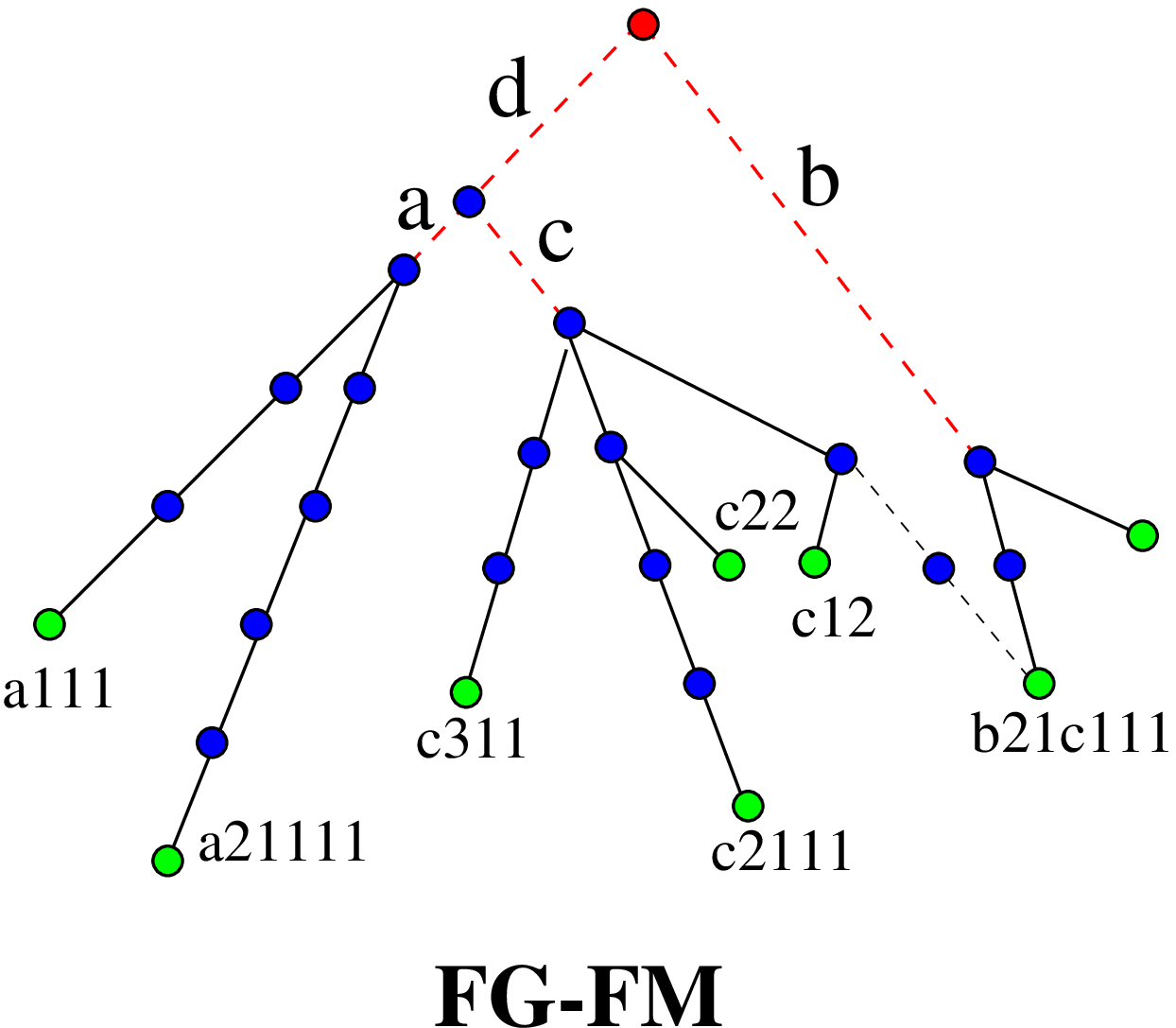}
  \end{minipage}
  \hfill
 \caption{ \label{Shad200} 
On the left we see the low noise $S_{200}(x)$  on chromosome 22 in our FG-FM family.  The partition of the samples responsible for the $S_M=1$ region is     $\{b_1,b_{12}c_{111}\}$ and this set's complement.  On the right we see a version of the FG-FM pedigree consistent with a disease locus on chromosome 22 as discussed in Section~\ref{Cake1}.  
    }
\end{figure}

 This method also applies to the FG-FM family, though in a second weaker form.  Once again we will explore the possibility of an IBD region in the $S_M=1$ candidate region.    In this case there is a unique partition that gives our candidate $S_M=1$ interval   and it is composed of the samples   $\{b_1,b_{12}c_{111}\}$ and this set's complement.  If we believe that an IBD region might be present, then   we would conclude that the true pedigree is more likely to look something like the pedigree in  Figure~\ref{Shad200},  with the relatively large number of non-founders  $d+b$  compared with the number of non-founders $a+c$.  With $d+b$  relatively large there are many chances for crossovers near the disease locus and hence the IBD region may be quite small.    To explore this possibility, we can look at an approximation with a better $Q$-Value, like $S_{50}(x)$ as in Figure~\ref{Shad50}.   Using  $S_{50}(x)$  we find a candidate IBD region.   The assumption  that $b+d$  is relatively large compared with $a+c$ makes   plausible the scenario for the IBD region's existence as pictured in the right half of Figure~\ref{Shad50}.   Furthermore, this IBD streak has a length of 60 markers  and  having a streak this long inside of our length 283 $S=1$ region by chance is unlikely.    Namely, in Section~\ref{Noise}, we find that  the probability  of a streak this long or longer  due to chance is less than 0.12.  While this argument is not as convincing as the earlier example with the FS-Z family (where the $S=1$ region was comprised of two partitions), this still gives us a good  first place to look for a disease allele.

A key aspect of this method that we still need to discuss, is how to choose $M$.   The ideas is to choose if possible an $M$ that simultaneously makes the chance of false positives and false negatives using the full {Shadow Method}   small.    In the next section, we estimate the false negative rate using the  {Shadow Method}  and in Section~\ref{Estimates} demonstrate how to   use the false negative rate to choose $M$.



\subsection{False negatives revisited \label{FN}} 

By applying the {Shadow Method} (and not simply attending to the regions where $S = 0$) 
 reduces the false negative rate.   We call this improved estimate of the false negative rate the $FN$-Value.      Notice,  $Q$ corresponds to  the false negative rate using just a streak analysis, while $FN$ corresponds to  the false negative rate using the full {Shadow Method}.   In Section~\ref{QVal}, we find that 
\[ FN  =  1 - \left(2+ \left( \frac{M B G}{N}\right)^2 - e^{-  \frac{M B G}{N}}\right) e^{- \frac{M B G}{N} }.  \] 
In Section~\ref{Estimates} we will quantify the extent to which this method enhances the use of 
$S$   via some examples.   This improvement in the false negative rate is the motivation behind the introduction of the full {Shadow Method} (as opposed to performing only a longest streak  analysis).   

\subsection{Estimates \label{Estimates}}

First let us review.  The following parameters will be considered:

\[ M =  \mbox{Streak length lacking obligate recombinations} \]
\[ B  = \mbox{Branches in collapsed pedigree} \] 
\[ N = \#(SNP markers) \]  
\[ D= \#(\mbox{Samples for which we have SNP data}) \] 
\[  p \approx P(\mbox{More Likely Allele}). \]
It is possible to assess from these parameters the   potential effectiveness  of the relevant {Shadow Method}.  For example, using  $M=200$ and $N= 100k$ we find for  our example pedigrees: 

$$\begin{array}{c}
\begin{array}{l|l|l|l|l}
 \; & P& FP& FN& Q\\
\hline
FS-Z & 1.4 & (10)^{-4} & 0.01 & 0.15 \\
\hline
FG-FM & 4\times (10)^{-4} & (10)^{-5} & 0.17 & 0.55 \\
\end{array}\cr
\;\cr
N=100k \mbox{ and } M =200
\end{array}$$

These values indicate  with $M=200$,   the FS-Z  pedigree would  be handled very nicely via  100k SNP marker sets since   both  $FN$ and $FP$  are reduced below $\frac{1}{100}$ (though the $P$-Value for this family is  weak and we would expect that we would need to  carefully try to  list all the IBD regions, see Section~\ref{PBig}.)  For the FG-FM family we see that  $M=200$ 
has a small $FP$ but a  rather large $FN$. 
If we were to try  $M=50$ we have the opposite problem

$$\begin{array}{c}
\begin{array}{l|l|l|l|l}
 \; & P& FP& FN& Q\\
\hline
FS-Z & 1.4 &114 & 6\times (10)^{-5} & 0.01 \\
\hline
FG-FM & 4\times (10)^{-4} & 65 & 0.002 & 0.08 \\
\end{array}\cr
\;\cr
N=100k \mbox{ and }  M =50
\end{array}$$

in which $FP$ is very large and  $FN$ is small.   
Exploring these values we see that we must make a compromise.  For example, for   $M=100$ we have:

$$\begin{array}{c}
\begin{array}{l|l|l|l|l}
 \; & P& FP& FN& Q\\
\hline
FS-Z & 1.4 & 1.2 & (10)^{-3} & 0.05 \\
\hline
FG-FM & 4\times (10)^{-4} & .32 & 0.025 & 0.23 \\
\end{array}\cr
\;\cr
N=100k \mbox{ and } M =100
\end{array}$$

If we don't wish to compromise we will need to use a denser mapping.   For example, using  500k  SNPs and    $M=200$ we have:

$$\begin{array}{c}
\begin{array}{l|l|l|l|l}
 \; & P& FP& FN& Q\\
\hline
FS-Z & 1.4 & (10)^{-3} & (10)^{-4} & 0.01 \\
\hline
FG-FM & 4\times (10)^{-4} & (10)^{-4} & (10)^{-3}& 0.05 \\
\end{array}\cr
\;\cr
N=500k \mbox{ and }  M =200
\end{array}$$

Hence we see in this case  the extra SNPs would really pay off.

 These estimates also give a sense of the future for SNP technology.  It is widely estimated that on average, two   genomes differ at 1 in 1000 nucleotides (i.e., approximately 3 million variants per genome). 
Hence, it is quite reasonable that we may find 5000k reasonably informative SNPs.  
 In this case a {\em Shandow } based approach applied to a
collapsed pedigree with 50 members, of which 10 are affected and sampled,
then using $M=190$, both $Q$ and $FP$ would be less than $1/500$.

\subsection{Assumptions and Caveats \label{Assump}}

Here we discuss the assumptions that underlie our analysis.   We  assume  that the markers occur randomly (with respect to morgan measure)
throughout the genome and that the  rates in the founder population of the
more common marker alleles behave as if they were randomly distributed
among the SNPs.    Violations of these assumptions will make some IBD regions easier to find and some harder.  Moreover, it is well known that such a random independent 
distribution is not going to be accurate SNP rates at which linkage
disequilibirum is observed (see \citet{Altshuler})  and   the SNPs in haplotypes contain less information do to the violations of independence.

Another simplifying assumption we make is that we can make a reasonable choice of a collapsed pedigree with a  common founder.
Of course on some scale, many ancient founders of all or most of the affected samples will exist.  However,  most such founders are too genetically distant to be picked up with our methods.  It is also much less likely that one of these alternate distant founders has introduced a disease allele into our population, at least for rare diseases caused by alleles of strong effect.

In general, the need to apply the full {Shadow Method}   will become less necessary to as the marker densities increase and the $Q$-Value shrinks. 
However  our estimation techniques rely on assumptions which are reasonable  for  the current SNP densities but may hamper the exploration of very large pedigrees with very dense SNPs. For example, this 500k and 5000K  estimates form the previous section  assume that the more common of the two SNP alleles occurs  on average no more than about  $85\%$ of the time in the founder population, which is true in our $10k$ and  $100k$ samples but may increase as SNP density increases hence increasing $FP$ (see Section~\ref{Noise}).

\onecolumn

\begin{figure}
 \begin{minipage}[t]{5cm}
 \includegraphics[width=0.85\textwidth]{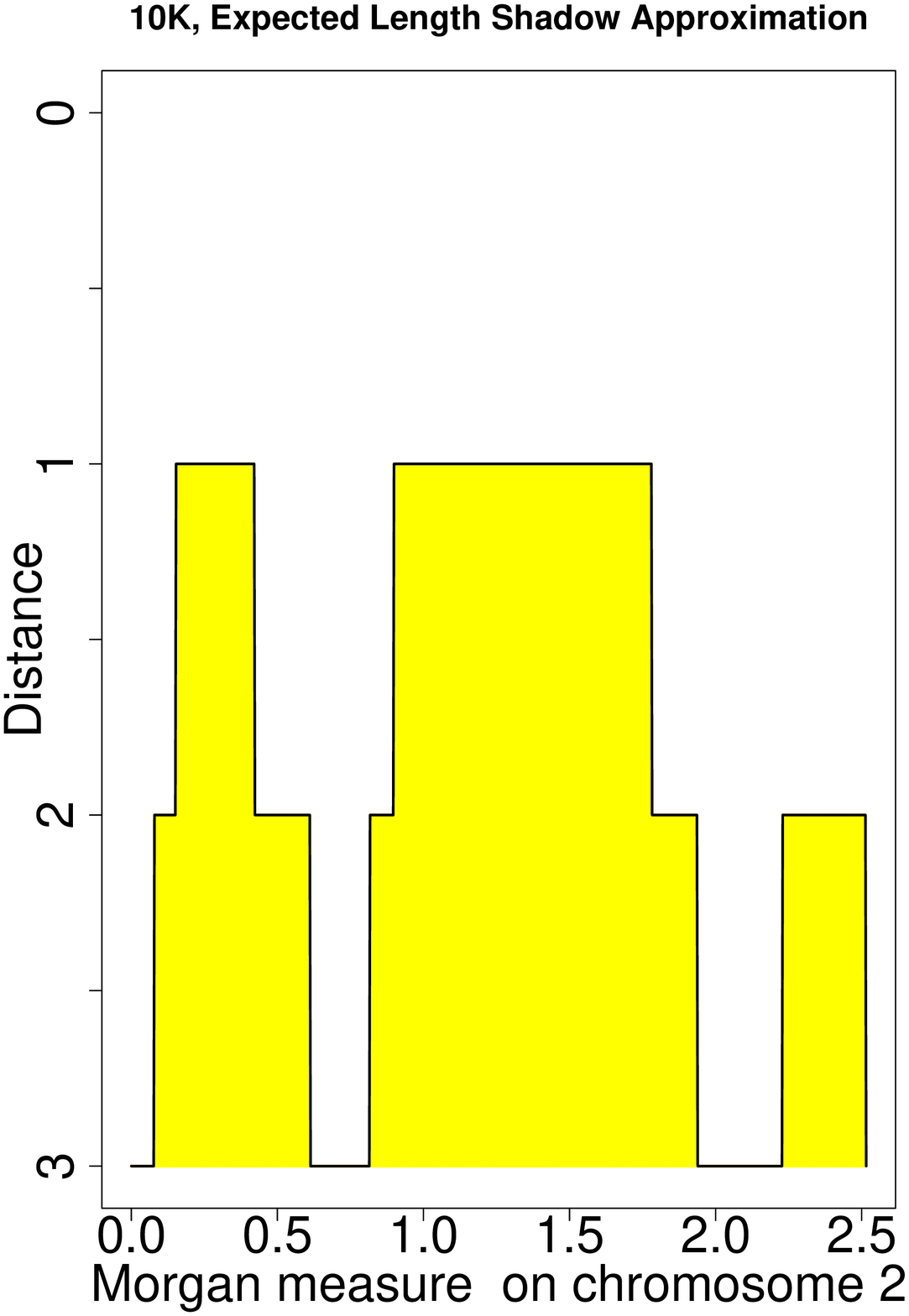}
 \end{minipage}
 \begin{minipage}[t]{12cm}
 \includegraphics[width=0.85 \textwidth]{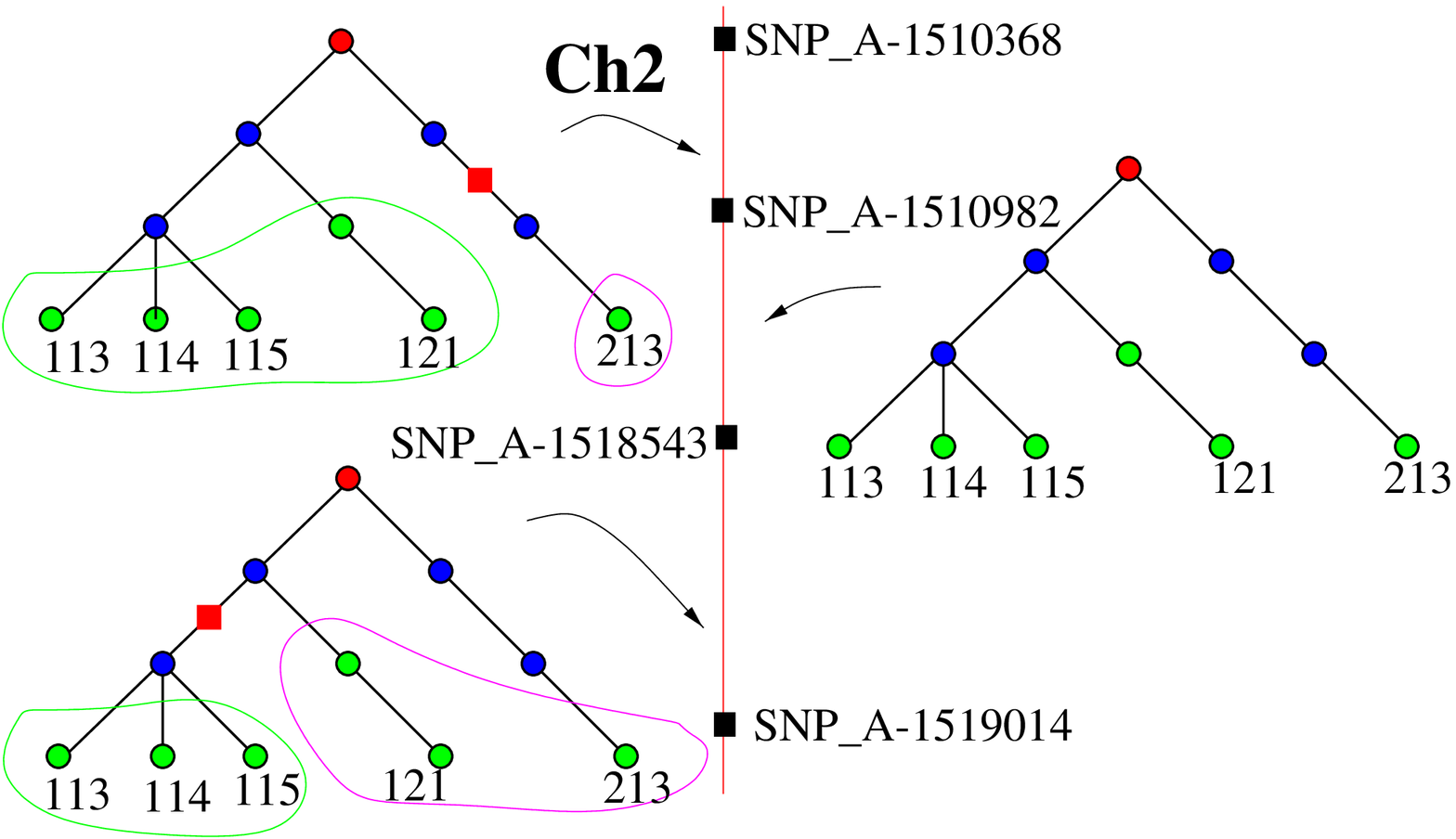}
 \end{minipage}
 \caption{ \label{Flank} 
 Here we see the next most promising candidate region in FS-Z, the  large $S=1$ region on chromosome 2.  In Section~\ref{Cake1}, we conclude the likely existence of an IBD region inside this $S=1$ region.  In the right half of this of this Figure we see  plausible positions in the pedigree  for the  obligate crossovers  (indicated  with yellow squares)  necessary to form the indicated partitions as determined by the {Shadow Method}.  
    }
\end{figure}

\begin{figure}
\hfill
 \begin{minipage}[t]{8cm}
 \includegraphics[width=0.85\textwidth]{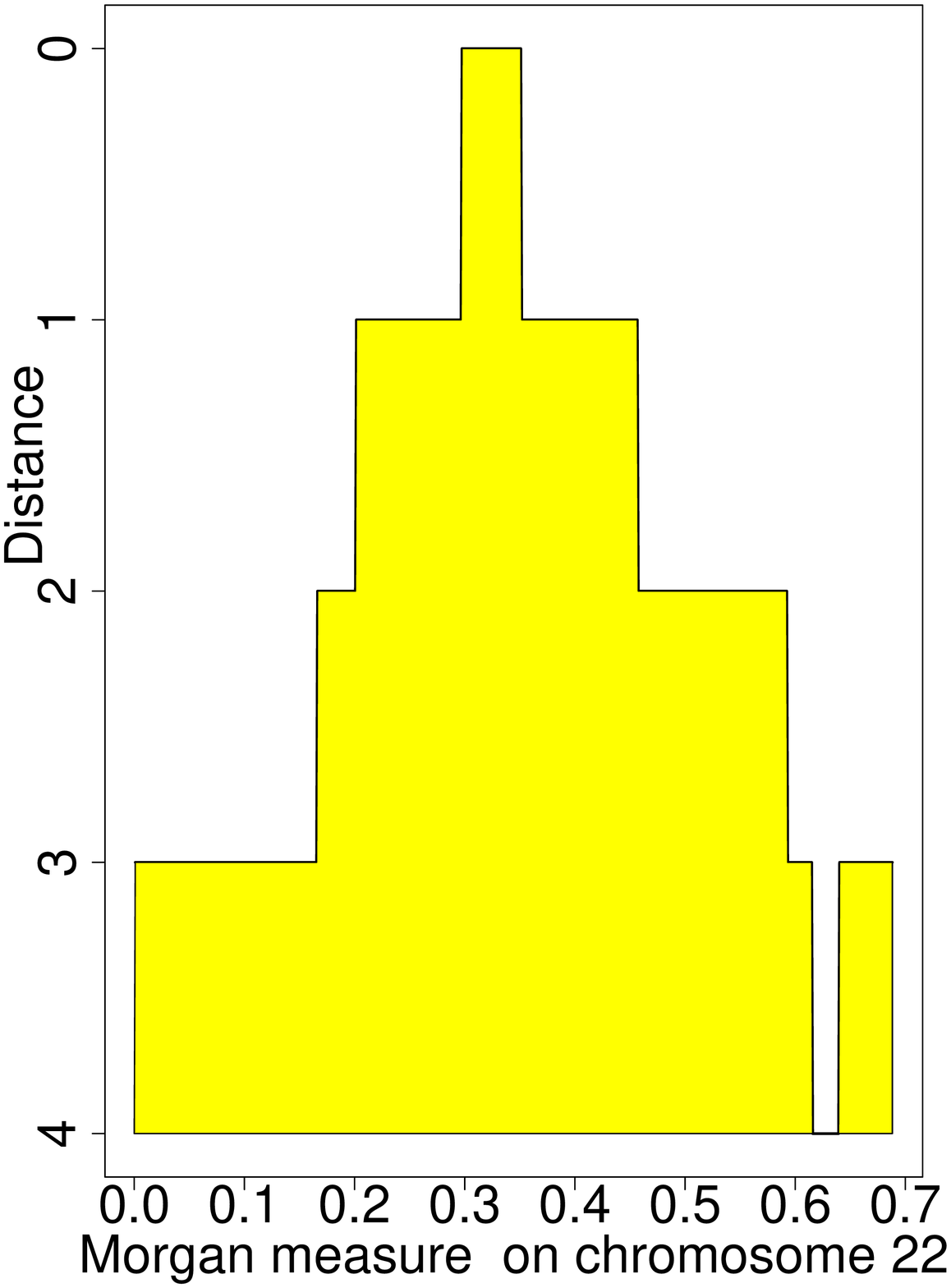}
   \end{minipage}
   \hfill
 \begin{minipage}[t]{5.5cm}
 \includegraphics[width=0.85 \textwidth]{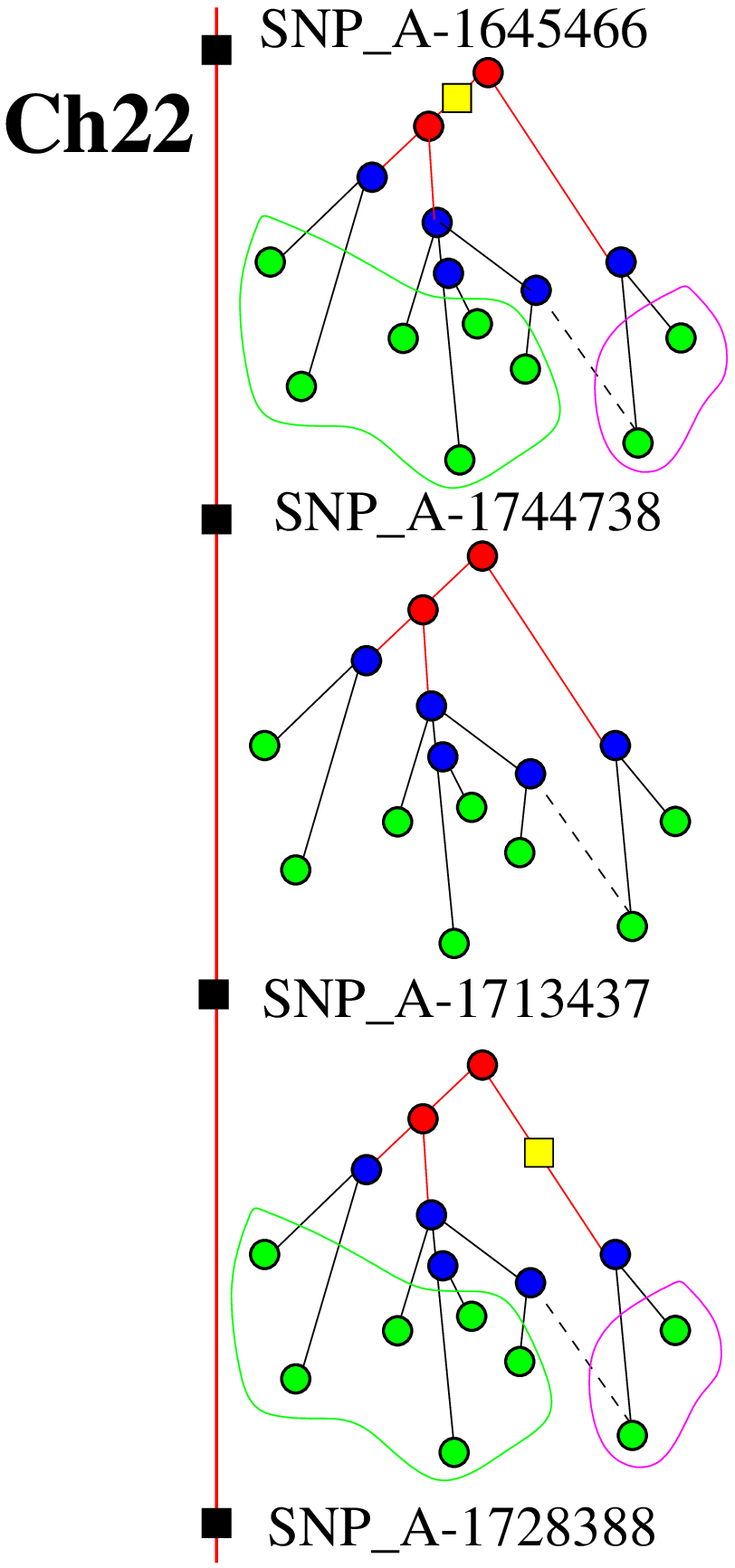}
 \end{minipage}
 \hfill
\caption{On the left  we see  the noisier  $S_{50}(x)$ that exposes  
our most likely candidate IBD region.     As in Figure~\ref{Flank}, on the right we explore the plausible IBD region. 
}\label{Shad50}  
\end{figure}

\begin{figure}
\epsscale{.25}
\plotone{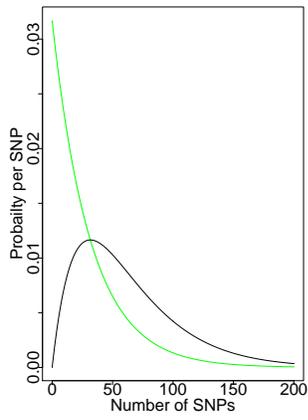}
\caption{ \label{Paradox}    ``The Disease Paradox'':   Assuming no a priori knowledge of the location of the  disease loci, we have that  the chance of the disease being located    in any given crossover interval is proportional to the length of the interval.  Hence the probability density function (pdf) of the length of the  interval containing the disease $f_{Disease}(l)$ satisfies   $f_{Disease}(l) \propto 
l f_{Chance}(l)$ where $f_{Chance}(l)$ is the  pdf of the of  length of a chance IBD region.     The green curve in this figure is the distribution of the  SNP length of  a chance IBD region, while black curve is the distribution of the SNP length of  a region which is  IBD because the disease is conditioned to be there. These curves were derived using  the independence model discussed in Section~\ref{QVal}. 
   }
\end{figure}

\twocolumn

\section{Computational Methods \label{Comp}}

The main purpose of this section is to discuss the complexity of the  algorithm \footnote{ The {\em complexity} of an algorithm is the number of arithmetic operations required.}  used to compute the {Shadow} and perform the {Shadow Method}.   To make this analysis we use the parameters reviewed in Section~\ref{Estimates} together with the definitions:

{\it 
$ T = \#$ of branches remaining upon removal of the non-genotyped pedigree members  (as in the pedigrees on the left hand side of Figure~\ref{Shad50})}

and

{\it 

  $H$ =  the maximal number of inconsistencies that we will be considering  in the computation of the  {Shadow}.  
  }

  For example, for the FS-Z family $T=7$, we choose $H=3$, and $D=6$ (recall $D$ is the number of samples). 
     
The algorithm  requires knowledge of the  {\em confidence call for a SNP}, and one must choose how to  throw away suspicious measurements.
This parameter is important since the {Shadow Method} is not robust under SNP miscalls.  (We used  parent/child  comparisons to help interpret this  error rate and found that a cutoff of $0.01$  using the Affymetrix  confidence call  works well.)

The analysis of the {Shadow} presented in the previous section made use of  an approximate pedigree.
This pedigree should be used only if there is a great deal
of    confidence that the disease allele is likely to be affecting the samples  via these known relationships.  The real power of the {Shadow Method}   is that it allows us to be more flexible if we are uncertain about the pedigree or of  the pedigree's role  in the spread of the disease.  Any likely pedigree can be used, but the complexity increases with each possibility.       Denote as $Ped_H$ the  collection of all partitions with $m(v) \leq H-1$ in the pedigree(s) of interest.    Then the complexity is    $O( N \left| Ped_H \right| )$.    If  there is just a single    pedigree to investigate then  we have the universal bound $\left| Ped_H \right|  \leq \sum_{k=0}^{H-1} {{T}\choose{k}}$.
 For a tree\footnote{We mean here "tree" in the graph theoretic sense
- that is a graph without loops} this bound is sharp, and for (the tree) FS-Z    $\left| Ped_3 \right| = 29$.    
 Notice if we have a list of candidate  collapsed pedigrees then $Ped_H$  is easy to construct by simply  adding in crossovers to the pedigrees and recording the resulting   partitions.
Typically, given a pedigree in which there is confidence both in the pedigree structure and clinical data, then this algorithm will work for very large pedigrees and number of samples  (certainly $T$ and $S$ both less than $22$ will work).

However, in practice the {Shadow  Method} will be most useful when   pedigree information is missing.  
  If we are completely open-minded about the pedigree structure, then   $\left| Ped_H \right|$ is less than or equal to the number of partitions of a set of $S$ elements into $H-1$ or fewer parts (and this number of parts determines the $m(v)$).  In other words, $\left| Ped_H \right| = \sum_{k=1}^{H-1} \left\{
 \begin{array}{ll}
 D \\
 k 
  \end{array}
   \right\}$
 where
  $\left\{
 \begin{array}{ll}
 D \\
 k 
  \end{array}
   \right\}$ 
   is a Stirling number of the second kind.  This sum grows exponentially and nearly at the rate $(H-1)^D$.  For the FG-FM analysis, we performed a  completely open-minded analysis  and choose $H=4$.  
   On our machines, we could not exceed  not exceed $S=11$ and $H=4$ with  100k data.
One important difference between this method and other forms of linkage analysis is that the size of the pedigree does not affect computational speed.  Rather the number of samples studies (irrespective of the structure of the pedigree) determines computational size and speed.
This will allow for the analysis for very large and complicated pedigrees.

 {\bf Comments About Controls:}
 This algorithm (and the {Shadow Method} itself) can be altered to incorporate controls.  
  For example:  Call a region ideally consistent with disease (ICD) if the affected samples are IBD in this region and the unaffected samples are not  related to  each other or the affecteds in this region. 
Then we can search for how far we are from an ICD region using the same exact techniques as that we designed to search for how far we are form an IBD region.  For example, the partition having a part for each unaffected and a part for all the affected   would now have $S=0$, and we would be estimating the distance from this situation.   be said that in this case, more inconsistencies should be
expected, since
the penetrance rate for a single allele might be very low. (Especially if the disease is recessive.  However, at a potentially recessive allele the algorithm can
be modified  to break a streak if an AB is observed, hence isolating the region around a recessive disease locus. Under the assumption that both mutant alleles are the same, which is of course  a very strong assumption.) To  optimally exploit  sibling and parental controls with such a streak analysis requires more work and a haplotyping version of the method, work we hope to describe this in future paper.

\section{Discussion} 

   We have described a simple method for identifying disease gene loci
in pedigrees using dense genetic data.  We believe this method has
several strengths.  In any family-based study designed to identify
loci harboring rare alleles of strong effect, the goal is to identify
a genetic locus (or loci) harboring alleles cosegregating with a
phenotype of interest.  The {Shadow} function defined here gives an
intuitive interpretation of dense genetic data.  At each point in the
genome, {Shadow} tells us how inconsistent that point is from being
located in a genetic region shared by a group of phenotypically
``affected''
individuals.  These $S=0$ regions are similar to regions
where the LOD score reaches its maximum attainable value.  However,
in contrast to a LOD score, the {Shadow} is not itself a likelihood
ratio.  Thus, for a family consisting of a single pair of affected
sibs, $S=0$ for half of the genome, and $S=1$ for the other half.  In
this method, statistical significance is  assessed separately.  We
assign to each value of $S$  a $P$-Value which describes the probability
of seeing this value by chance.  We also generate $FP$ and $FN$  values, so that we can assess the chances of a false positive and false negative using this method.    
In turn, these estimates allow us to make a priori estimates of an appropriate choice 
of the key parameter $M$, the length of a streak of markers lacking obligate recombination events. 

       In addition, the {Shadow Method} helps us identify the cause of
deviations from $S=0$ regions.  For example, in a genome-wide analysis,
we may find no $S=0$ region, but a small number of $S=1$ regions.  We
can  specifically examine the nature of the one inconsistency in each
such region to help us evaluate the plausibility that a phenotype-causing allele is in fact present.

       This method has limitations.  There is certainly no practical reason
to use {Shadow} to analyze a pedigree of the size of FS-Z where a
standard linkage analysis with a map of only moderate density will
work well.   While at the  currently routinely available SNP map densities (such as Affymetrix 10k and 100k SNPChips) the assumptions we use in our analysis appear reasonable, we must hope that the nature and quality of SNPs does not change in significant ways as densities increase or our estimation method will fail to make good sense and will need to be modified.

       As noted, we plan to develop further refinements of this methodology
allowing the incorporation of a greater fraction of the available
genetic information as well as data from controls and
unaffected family members.
  However, in its present form, we believe
{Shadow} will have immediate value in the analysis of genetic data in
complex family studies in which traditional linkage analysis
calculations are problematic.



\acknowledgments
  
  
 \appendix

\section{$P$-Value \label{PVal}}

Here we explain how to approximate  the  required  $P$-Values.  
We carefully justify our computation in the basic  case where $P$ is the probability of an IBD region ($S=0$ region) and the approximate pedigree is a tree; we then explain how modify the answer for  other values of $S$ and more complicated pedigrees.   The first observation is that this $P$-Value is bounded by the expected number of IBD regions, and it is this quantity that we compute.   

For each pair  of spousal founders there are four chromosomes which could 
be responsible  for a given IBD region.  We fix one of these four possibilities  for  the $i^{th}$  chromosome and call a region of the samples  IBD relative to  it  ${ \mbox{IBD}}_i$. Recall $ch_i = E(C_i)$ where $C_i$ is the  total number of crossovers during a meiosis process on the $i^{th}$ chromosome.  We have
\[ \begin{array}{lll}
  E(|\mbox{ IBDs}|) &  =  &  4   E(\sum_{i=1}^{22}  \#({ \mbox{IBD}}_i{\mbox{s}}) )  \\
 &  = &  4 \sum_{i=1}^{22}   E(E(\#({ \mbox{IBD}}_i{\mbox{s}}) \mid C_i = N  )) \\
 &   = &  4 \sum_{i=1}^{22} E(\frac{N+1}{2^B}) \\
 &  =  &  4 \sum_{i=1}^{22} \frac{B ch_i+1}{2^B}.
   \end{array}
 \] 

  Notice this estimate of the $P$-Value   is exponentially decreasing.  In particular, if the collapsed pedigree is a tree then  for more than 16 branches  the chance of a chance IBD is less than 5 percent, and if the number of branches is greater than 20 then the $P$-Value of an chance IBD region  is less than 1 in 500.

 For a non-tree the numerator of $\frac{1}{2^B}$ would become the number of collections of  crossover events  that still leave our  samples IBD. For the FG-FM pedigree, we have one loop and find that    number of collections of  crossover events  that still leave our  samples IBD  equals $2^2+2^2-1$. So the expected number of IBD regions is less than $\frac{1}{2500}$ and hence the $P$-Value associated to an IBD candidate is  bounded by   $\frac{1}{2500}$.    For a general $S=k$, we must list all the partitions that are consistent with $k$ or fewer obligate crossovers and then count all the collections of  crossover events that can result in such  partitions.  We find  that the probability of an $S=1$ region in our FG-FM family is less than $\frac{1}{25}$.

\subsection{When the  $P$-Value is high \label{PBig}}

Using the computation in the previous section, in FS-Z we find that we expect 1.4 IBD regions from this pedigree other than the one due to the disease.  This explains  why we should not be surprised to find at least two IBD regions (as we see in Figures   \ref{LodComp} and \ref{Flank}).  In general, it important to list all the candidate regions when the $P$-Value is not small.  For example, under the assumption that we have no information about the location of our disease loci before the experiment, we have the following  theorem:

 \medskip
{\bf Key Theorem:} Assuming  the disease is in  $D=\{x \mid S = k\}$, 
 the probability that a given interval in $D$ contains the disease marker is proportional to that region's length in $D$.
\medskip

For example, using the  {Shadow Method}  we find three good candidate IBD region in the FS-Z family (the ones on chromosomes 11 and 2 and another on chromosome 16) with morgan lengths roughly $0.2$, $0.05$  and $0.05$.  So assuming the disease loci was in an IBD region, this key theorem tells us that we should have assigned an a priori  probability of roughly $2/3$ of the disease loci  being    in chromosome 11 region (where it turned out to actually be). 

\section{False Negatives \label{QVal}}

Here we approximate  the $Q$ and the $FN$. 
  To compute the $Q$, we first notice that 
the end points of the interval corresponding to the interval with a disease locus  are dictated by crossovers in the collapsed pedigree.   
The length of a randomly selected such interval will have a length distribution corresponding to the   distribution of the length of a chance IBD region, and we denote the probability density function (pdf) of this length as $f_{Chance}(l)$. As it turns out, the disease is more likely to be in a longer interval.  This  fact sometimes goes by  the name of the ``Bus Paradox'', which for the purposes (and context) of this paper, we  will call rename the ``Disease Paradox''. This is illustrated in Figure~\ref{Paradox}, where we find the pdf of the interval containing a disease loci is  $f_{Disease}(l) \propto  l f_{Chance}(l)$.


To determine $f_{Chance}(l)$  requires a choice of model for the meiosis process. 
We use the standard independence assumption that underlies the linkage analysis 
approach as developed in \citet{Lander} and call this model the {\it independence model}.     Here is a brief review.   We can view each chromosome as an interval with subintervals  each associated to an inheritance vector where neighboring inheritance vectors   differ by exactly one change to the vector.  We call such an interval a {\it crossover interval} when we restrict our attention to the collapse pedigree (with respect to  any one of the four founding chromosomes).       We need to decide how to choose the endpoints of these subintervals.    There is a natural measure on each chromosome which assigns to each interval  the expected number of cuts during a meiosis process, called the {\em Morgan measure}.  In the case of multiple  cuts during  meiosis, the positions  are not independently chosen with respect to the Morgan measure (this is due to interference),   but  if we view the meiosis process associated to distinct  individuals in our  pedigree  as independent and note  that the expected number of cuts per individual is small,  then   a Poisson process should  give an excellent approximation when examining  even a moderate size pedigree.   This approximation is  equivalent to the well studied Markov assumption as utilized in most forms of linkage analysis  and  as developed in  (\citet{Lander}) and in (\citet{Kruglyak}).   This model is not directly utilized in the formulation of our algorithm, but only utilized in order to analyze the results  and it is also how we simulated  the genetic process. (We assumed independent founders and choose the cuts via  this Poisson process.)

 Under our Poisson assumption   $f_{Chance}(l) = B e^{-B l}$, hence  $f_{Disease}(l) = B^2 l e^{-B l}$.  To estimate the $Q$-Value, first note that $M$ SNPs  corresponds roughly to $m = \frac{G M}{N}$ morgans, and hence under these assumptions  we can approximate the   the $Q$-Value via 
\[  \begin{array}{lll}
 Q  & = &   B^2  \int_{0}^m l e^{-B l} dl  \\
   & = &    \int_{0}^{Bm } l e^{- l} dl  \\
  & = & 1- (1+B m) e^{-B m}.
\end{array} \]

To approximate $FN$ we can look one layer down in the tree.  To do so, let $L$ denote the length of the region to the IBD region's left and $R$ the length of the region to its right.   We have
  \[ \begin{array}{lll}
 FN  & = &  \int_{0}^m   P(R \leq (m-l) \mbox{ and } L \leq (m-l))  f_{Disease}(l) dl  \\
 & = &  \int_{0}^m   P(R \leq (m-l)) P( L \leq (m-l))  f_{Disease}(l) dl  \\
  & =&  \int_{0}^m  (1-e^{-B (m-l)})^2  B^2 l e^{-Bl}  dl   \\
   & =& 1 - (2+ (mB)^2 - e^{- mB}) e^{-m B}. 
  \end{array}
 \]

\section{False Positives  \label{Noise}}

To explore noise we need to articulate a model of the SNPs themselves.  We let  $\mbox{SNP}_i$ denote the value of the $i$th  SNP.   Each  $\mbox{SNP}_i$  comes in one of two flavors, $A$ or $B$.  It is perhaps more useful to think of them labeled instead as  $Less$ and $More$, representing  the 
{\it less} and {\it more} common alleles.  To model the distribution of the SNPs  we would like to assign a value of $More$  with a probability that approximates the rate at which  $More$  would occur in  a population of founders. Let us call this probability $P(\mbox{SNP}_i = More)$.    Note that   if we intend to use a parametric maximum likelihood method (as in (\citet{Kruglyak})), then it would be wise to carefully explore this distribution.  However, for our purposes we feel some simplifying assumptions are reasonable, namely that the population from which founders are drawn is large enough so that the $\mbox{SNP}_i$  are independent, and that $p=P(\mbox{SNP}_i = More)$ is independent of $i$.
   We also used these assumptions when simulating of SNPs.   We acknowledge that these are  {\bf very} serious assumptions and will become falser and falser for denser and desner maps, as dicussed in Section~\ref{Assump}.    

As introduced in Section~\ref{Noise1}, Noise  is comprised  of streaks of data accidentally consistent with a given partition making $S_M(x) > S(x)$.  
Hence, we will define a {\it unit of noise} a  to be a streak of length greater or equal to $M$ where  $S_M(x) > S(x)$ throughout this streak.  We let {\it Noise} in a region be the total number of such units of noise in that region.

  Let us start with an example of estimating noise from the FG-FM family  by examining carefully our $S=1$ region on chromosome 22 in the FG-FM family.  Here we have an interval of length $L=283$ markers consistent with a partition of our 8 samples with two parts, one of size 2 and the other size 6.  Under our simplifying assumptions, we claim that the expect noise in the  the $M=50$ {Shadow}
 is  given by  
  \[ E(\mbox{Noise}) \leq L p_n (1-p_n)^M, \] 
where $p_n = p q ((1-p^6)(1-q^{2})+(1-q^6)(1-p^{2}))$.

{\bf Proof:}
The observation is simple.  We assign the position in our length $L$ region a value of $1$ if it starts a streak of length greater than or equal to $M$ and a $0$ otherswise and denote the quantity as $Pos_i$. 
Each $Pos_i$ is version of the random variable $Pos$ which is equal to 1 if M+1 flips of the coin are such that
the first result is a tail, and the next M give heads, this with a
probability of tails equal to $p_n$.
Hence $E(Pos) =  L p_n (1-p_n)^M$.
 
\[ E(\mbox{Noise}) = E(\sum_{i=1}^{L-M} Pos_i) =  \sum_{i=1}^{L-M-1} E( Pos_i) \leq L E(Pos)  = L  p_n (1-p_n)^M   \]

Now we need to estimate the probability of tails. In order to have a tails outcome, we need  the  chromosomes that are IBD for each part $Allele(\mbox{Part 1}) \neq Allele(\mbox{Part 2})$ and that at
least one of the other  founder chromosomes in each part takes on the same allele value as the chromosome that is IBD for this part. 
Hence 
\[ p_n  = p (1-p^6) q (1-q^{2})+ q (1-q^6) p (1-p^{2})) \] 
as claimed.

\hfill { \bf QED} \hspace{.1in}

Of course to  actually compute it we need an approximation of $p$. To do so, we first note that the probability that the alleles
are different,  $P(AB)$, equals 
$2 p(1-p)$.   We can use our data to approximate $P(AB)$ and solve this quadratic to find $p \approx 0.84$.  (This corresponds to a maximum likelihood estimate of the  parameter.)

If we use the whole genome as our region then  $FN < E(Noise)$ and it is this relationship we will use to bound $FN$.  To give a nice bound we use the special case of a tree, though only role this plays is in insuring that the chance  of an accidental consistent set of markers when $S<1$ is less than the this chance when $S=1$. Hence we can use the  $S=1$ case with a unique corresponding partition to bound this probability.  So we can assume there are two parts  in our partition of our  $D$ samples and hence the probability of success at any give point is  bounded above and below by

\[ p_{max} =  \max \{  p q ((1-p^{D-j})(1-q^{j})+(1-q^{D-j})(1-p^{j}))  \mid 1 \leq j  \leq \lfloor D/2 \rfloor  \}  \] 
\[ p_{min} =  \min \{  p q ((1-p^{D-j})(1-q^{j})+(1-q^{D-j})(1-p^{j}))  \mid 1 \leq j  \leq \lfloor D/2 \rfloor   \}  \] 

and the same argument as above tells us that 
\[ FP < E(\mbox{IBD Noise}) \leq    N  p_{max} (1-p_{min})^M, \] 
and hence will give us a sense for the expected noise.

In general, such estimates tells us  that if $N$ is big enough we do not need to be very careful in analyzing our data and the {Shadow Method} will work great.  For example letting $M= \sqrt{N}$ we can see that as SNP density gets thicker the percentage of the genome where the $S_M$ and $S$ disagree quickly goes to zero as $N$ goes infinity.  However as  discussed in Section~\ref{Assump}, as the marker density  increases  the assumptions that underlie our estimate will be become less and less realistic (especially the independence of the markers),  and caution is required. 

\section{Key Theorem}\label{key}

{\bf Set Up:} Let  $D$ be a set, let
$rp$ be a process that selects a random point from $D$,
and let $RS$ be a process that selects a random subset of  $D$.

{\bf For Simplicity:} Assume $D$ is finite and that   $P(x \in  RS) \neq 0$ for all $x$ (like the real human genome).


{\bf Definition:} Let $(RS \mid x \in RS)$ denote the result of the process conditioned to contain $x$. 


{\bf Lemma:}   Upon witnessing  $E = (RS \mid rp \in RS)$  we have 
\[ P(rp =x \mid (RS \mid rp \in RS)=E)  \sim  \frac{P(rp=x)}{P(x \in RS)}     \chi_E(x) , \]  
where  $\chi_E $ is the indicator function on $E$.

{\bf Proof:}  
Recall  {\it Bayes Theorem}  
 \[ P(A \mid B) = P(B \mid A) \frac{P(A)}{P(B)} \]  
   and notice  
\[ P((RS \mid x \in RS)=E) = P( RS=E \mid  x \in RS).  \]

 
 From these observations, we have that $P(rp =x \mid (RS \mid rp \in RS)=E)$ 

\begin{eqnarray*}
  &  = &  P( (RS \mid rp \in RS)=E \mid  rp =x  )  \frac{P(rp =x)}{P((RS \mid rp \in RS)=E)}  \\ 
& = &  P( (RS \mid x \in RS)=E )  \frac{P(rp =x)}{P((RS \mid rp \in RS)=E)} \\
& = &  P( RS = E  \mid x \in RS)  \frac{P(rp =x)}{P((RS \mid rp \in RS)=E)}  \\
& = &  P( x \in RS  \mid  RS = E  )   \frac{P(RS =E)}{P(x \in RS)}     \frac{P(rp =x)}{P((RS \mid rp \in RS)=E)}   \\
& = &  \chi_E(x)  \left(  \frac{P(RS =E)}{P((RS \mid rp \in RS)=E) }  \right) \left(    \frac{P(rp =x)}{P(x \in RS)}   \right)  
\end{eqnarray*}

as asserted.

\hfill {\bf QED}  \hspace{.1in}

{\bf Comment:} This lemma   captures the intuitive fact that  if a point is relatively unlikely to be in $RS$ but turns up in $E$,  then this point is  more likely to be the point upon which $RS$ was  conditioned. 
This could be useful in situations in which there is a great deal of prior
 information regarding the disease loci. 
However when applying this lemma to derive the key theorem  we assume that the disease's  location is a priori totally unknown  (so $P(rp=x)$ is independent of $x$) and that make the Medelian assumption ($P(x \in RS) $ is independent of $x$).

\clearpage

\end{document}